\documentclass[journal]{IEEEtran}
\usepackage{cite}
\usepackage{amsmath,amssymb,amsfonts}
\usepackage{algorithmic}
\usepackage{graphicx}
\usepackage{textcomp}
\usepackage{xcolor}
\usepackage{cuted}
\usepackage{epstopdf}
\usepackage{stfloats}
\usepackage{algorithm}
\usepackage{enumerate}
\usepackage{comment}
\usepackage{tikz}
\usepackage[caption=false,font=small,labelfont=sf,textfont=sf]{subfig}

\begin{document}

\title{Low-Altitude ISAC with Rotatable \\ Active and Passive Arrays}
\author{Ziyuan~Zheng, Qingqing~Wu, Yanze Zhu, Honghao Wang, Ying Gao, Wen~Chen, and Jian~Xiong  
\vspace{-18pt}
\thanks{Z. Zheng, Q. Wu, Y. Zhu, H. Wang, Y. Gao, W. Chen, and J. Xiong are with the Department of Electronic Engineering, Shanghai Jiao Tong University, 200240, China  (e-mail: \{zhengziyuan2024, qingqingwu, yanzezhu, hhwang, yinggao, wenchen, xjarrow\}@sjtu.edu.cn). \textit{(Corresponding author: Qingqing Wu.)}}
}

\markboth{}%
{Shell \MakeLowercase{\textit{et al.}}: Bare Demo of IEEEtran.cls for IEEE Journals}

\maketitle


\begin{abstract}
This paper investigates a low-altitude integrated sensing and communication (ISAC) system that leverages cooperatively rotatable active and passive arrays. We consider a downlink scenario in which a base station (BS) equipped with a three-dimensional (3D) rotatable active array serves multiple users while sensing low-altitude targets in monostatic mode, assisted by a rotatable reconfigurable intelligent surface (RIS). To capture the impact of array orientations, we develop a rotation-aware geometry-based multipath model in which 3D rotations jointly affect the phases of the steering vector via rotated element coordinates, and the direction-dependent element and antenna radiation gains via a practical directivity model. On this basis, we formulate a joint design problem that maximizes a weighted-sum utility of the downlink sum rate and the sensing beampattern similarity, where the latter is quantified by a normalized mean-squared error (NMSE). We introduce a scaling parameter and update it in closed form once per outer alternating-optimization (AO) iteration. The resulting non-convex problem is tackled via an efficient AO framework that alternately updates the BS precoder, RIS phase shifts, and array rotation angles, using a quadratic transform and majorization-minimization-based closed-form updates, Riemannian conjugate gradient on the unit-modulus manifold, and projected gradient ascent under rotation box constraints, respectively. Numerical results demonstrate that the proposed jointly rotatable arrays at the BS and RIS achieve a favorable rate-sensing trade-off and performance gains compared with fixed or partially rotatable baselines.
\end{abstract}
\vspace{-6pt}

\IEEEpeerreviewmaketitle

\vspace{-6pt}
\section{Introduction}
\vspace{-3pt}
Low-altitude wireless networks (LAWNs) are emerging as a key component of next-generation intelligent infrastructures, interconnecting uncrewed aerial vehicles (UAVs), low-altitude platform stations, aerial robots, and connected ground terminals in the three-dimensional (3D) airspace below a few kilometers \cite{ref1,ref2,ref3,ref4}. Typical LAWN applications, such as aerial surveillance, disaster response, smart-city sensing, and low-altitude traffic management, rely not only on reliable broadband connectivity but also on accurate situational awareness of dynamic low-altitude objects. This naturally calls for integrated sensing and communication (ISAC) designs that can jointly support data transmission and high-resolution sensing within complex, rapidly changing airspace environments \cite{ref5,ref6}. Compared with conventional terrestrial cellular systems, LAWNs must cope with rapidly varying spatial geometries and require shaping radio signals to illuminate aerial targets while meeting stringent size–weight–power constraints and capturing their reflections, which make efficient ISAC-oriented signal processing particularly challenging. 

Multiple-input multiple-output (MIMO) technology and advanced precoding have long been recognized as key tools to exploit spatial degrees of freedom (DoF) in wireless systems \cite{ref7,ref8}. MIMO systems provide significant gains in beamforming, spatial multiplexing, and diversity by leveraging independent or quasi-independent channel fading, thereby significantly improving spectral efficiency and link reliability. However, in classical designs, antenna arrays are fixed in position and orientation, limiting spatial control to electronic beamforming. For LAWNs, where users and targets are distributed in full 3D and may move at low altitude with complex trajectories, purely electronic beamforming is often not sufficient, as system performance becomes strongly dependent on the relative 3D geometry among the base station (BS), scatterers, and low-altitude objects. This motivates the exploitation of geometry-aware architectures that can reshape both the propagation environment and the array response itself for LAWN-specific ISAC.

Reconfigurable intelligent surfaces (RISs) offer a promising approach to tailor the wireless environment via nearly passive phase-shift elements that reflect incident signals toward desired directions \cite{ref9,ref10,ref11,ref35,ref38}. RIS-aided systems have been extensively investigated for coverage enhancement, interference mitigation, and energy-efficient MIMO operation. More recently, in addition to communication, RIS has been shown to improve wireless sensing, including target detection and localization, by redirecting illumination toward non-line-of-sight (NLoS) targets or acting as a passive reflector from favorable angles, thereby enriching spatial diversity \cite{ref12,ref13,ref14}. In parallel, movable antennas (MA) has recently been proposed to further enhance spatial flexibility by allowing antenna elements or arrays to change their positions within a small region \cite{ref15,ref16,ref17,ref18}, enabling substantial gains in diversity, multiplexing, and interference management \cite{ref19}, and have been extended to multicast scenarios \cite{ref20}, as well as to two-timescale designs exploiting statistical channel state information (CSI) \cite{ref21}. These and other related works on MA-enabled systems demonstrate that adding mechanical DoF to electronic beamforming can significantly improve performance \cite{ref36,ref37,ref39,ref40}.

However, most existing MA- or RIS-related works treat antenna movement or passive phase tuning in isolation \cite{ref22}, and their ISAC designs for LAWN-like geometries are much less explored. Existing ISAC designs typically assume fixed arrays and do not exploit low-altitude 3D geometry control to shape the transmit beampattern and the multiuser downlink channels simultaneously \cite{ref23}. Furthermore, the mechanical flexibility in the MA literature is typically modeled as small-scale position changes within a local region. In contrast, realistic low-altitude platforms, e.g., UAV-mounted arrays or rooftop BS panels, can be designed to support rotation of the entire active or passive array via low-cost mechanical actuators \cite{ref24,ref32,ref33}. Such rotatable arrays (RA) can jointly adjust their boresight directions, element coordinates, and directional gains, offering a distinct and practically relevant form of geometry control \cite{ref25,ref41}. In summary, RIS-aided ISAC designs that consider rotatable passive arrays and their coupling with active-array rotations at the BS have yet to be systematically studied, especially under explicit constraints on the similarity of sensing beampatterns. As a result, the fundamental question of how to jointly design rotatable active and passive arrays, RIS phase shifts, and multiuser transmit beamforming for low-altitude ISAC remains open \cite{ref34}.

Motivated by these observations, this paper investigates a low-altitude ISAC architecture with cooperatively rotatable active and passive arrays. We consider a downlink LAWN scenario in which a BS equipped with an active rotatable array simultaneously serves multiple single-antenna ground users and senses low-altitude targets in monostatic mode, assisted by a rotatable RIS, termed a passive array. Both arrays can rotate in 3D, and the resulting orientations jointly affect the element coordinates and equivalent phases in the steering vectors and direction-dependent array gains through a practical radiation model. Based on the rotation-aware channel model, we formulate a joint design problem on the BS precoder, the RIS phase shift vector, and the rotation angles of arrays, which maximizes a weighted-sum ISAC utility of the downlink sum rate and the sensing beampattern similarity, subject to the BS transmit-power budget, RIS unit-modulus constraints, and mechanical rotation limits at both the BS and the RIS.
The main contributions of this work are summarized as follows
\vspace{-3pt}
\begin{itemize}
\item We propose a LAWN-oriented ISAC architecture where a rotatable active BS array cooperates with a rotatable passive RIS array. We develop a rotation-aware 3D multipath channel model that captures both phase shifts of steering vectors induced from rotated antenna and element positions and directional path gains induced from direction-dependent antenna and element radiation patterns. We formulate an optimization problem that maximizes the weighted-sum utility on communication sum rate and sensing beampattern similarity, which is characterized by normalized mean-squared error (NMSE) between the generated and desired beampattern, by jointly designing BS transmit beamforming, RIS phase shifts, and active and passive array rotations, subjected to power budget on the BS, unit-modulus constraint on the RIS, and mechanical rotation bounds on both RAs.

\item We develop an alternating optimization (AO) framework that iteratively updates the BS precoder, RIS phase shifts, and array rotations. For the BS precoding block, we combine the quadratic transform (QT) and the Lagrangian dual transform (LDT) with a majorization–minimization (MM) step to obtain a tractable quadratic-constrained quadratic program (QCQP) subproblem that admits a closed-form water-filling-like update. For the RIS block, we exploit the complex-circle manifold structure and design an efficient Riemannian conjugate gradient (RCG) algorithm. For the rotation block, we derive closed-form gradients of both SINR and beampattern NMSE with respect to the BS and RIS Euler angles and employ a projected gradient descent (PGD) method with Barzilai–Borwein initialization and Armijo backtracking.

\item We conduct extensive simulations using LAWN-inspired 3D geometries and compare the proposed design with several baselines, including only BS rotation, only RIS rotation, fully fixed arrays, and non-RIS configurations. The results show that the jointly rotatable BS and RIS architecture provides substantial gains in the weighted ISAC utility of the communication rate and beampattern NMSE across varying transmit powers, antenna array sizes, and user densities. The gains are further amplified by practical directional radiation patterns, highlighting the importance of aligning the array boresights and main lobes with both the user and target directions in low-altitude 3D space. By sweeping the trade-off weight, we further characterize the Pareto frontier between the sum rate and the NMSE, thereby quantifying how joint array rotations expand the ISAC trade-off boundary.
\end{itemize}
The rest of the paper is structured as follows: Section II introduces the system model and problem formulation. Section III presents the proposed algorithm. Section IV presents the numerical results, and Section V concludes the paper.

\textit{Notations:} Scalars are denoted by italic letters, and vectors and matrices are denoted by bold-face lower and upper-case letters, respectively. $\mathbb{R}^{x}$ and $\mathbb{R}^{x\times y}$ denote the space of real vectors or matrices, and $\mathbb{C}^{x}$ and $\mathbb{C}^{x\times y}$ denote the space of complex vectors or matrices, respectively. $\left| x \right|$ denotes the modulus of a complex-valued scalar $x$. For a vector $\boldsymbol{x}$, $\left\| \boldsymbol{x} \right\|$ denotes its Euclidean norm and $\boldsymbol{x}^*$ denotes its conjugate. The distribution of a circularly symmetric complex Gaussian random vector with mean vector $\boldsymbol{x}$ and covariance matrix $\boldsymbol{\varSigma }$ is denoted by $\mathcal{CN} \left( \boldsymbol{x},\boldsymbol{\varSigma } \right)$. The Euclidean gradient of a scalar function $f(\boldsymbol{x})$ with a vector variable $\boldsymbol{x}$ is denoted by $\nabla f(\boldsymbol{x})$. For a square matrix $\boldsymbol{S}$, $\mathrm{Tr}\left( \boldsymbol{S} \right)$ denote its trace. For any general matrix, $\boldsymbol{M}^H$ and $\left[ \boldsymbol{M} \right] _{ij}$ denote its conjugate transpose and $(i,j)$th element,respectively. $\boldsymbol{I}$ denotes an identity matrix, respectively. $\mathbb{E}[\cdot]$ stands for the expectation operation.

\setlength{\abovecaptionskip}{0pt}
\begin{figure}
\centering
\includegraphics[width=3.4in]{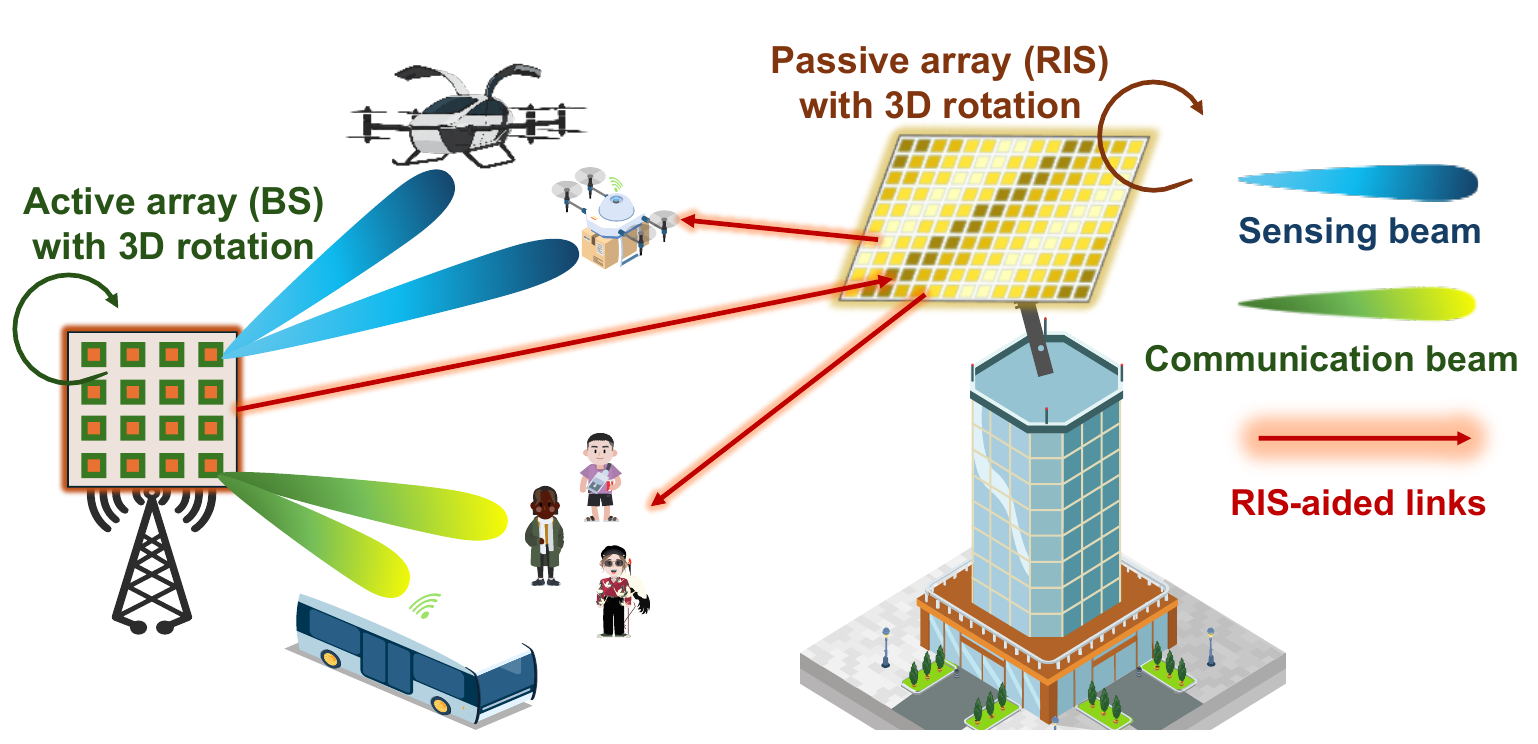}
\captionsetup{font=small}
\caption{A low-altitude ISAC system with rotatable active and passive arrays.} 
\label{fig1}
\vspace{-18pt}
\end{figure}

\vspace{-6pt}
\section{System Model and Problem Formulation}

We consider a downlink RA-enabled ISAC system, as shown in Fig. \ref{fig1}, where a BS uses an active rotatable uniform planar array (UPA) with $M$ transmit antennas to serve $K$ single-antenna users while sensing multiple targets in monostatic mode; at the same time, a passive rotatable RIS arranged as a UPA with $N$ reflecting elements assists both tasks. In addition to active transmit beamforming at the BS and passive phase shifts at the RIS, both arrays can rotate in 3D. These rotations steer the array boresights and reshape the array response to improve or balance the performance between communication and sensing.

\subsection{Array Geometry and Rotations}

We adopt a global right-handed Cartesian coordinate system (CCS) $\mathcal{G} = (O,\mathrm{x},\mathrm{y},\mathrm{z})$. The BS employs a co-located transmit/receive UPA with size $M=M_{\text{col}}\times M_{\text{row}}$ of $M_{\text{col}}$ columns and $M_{\text{row}}$ rows, and the RIS is also a UPA with size $N=N_{\text{col}}\times N_{\text{row}}$ of $N_{\text{col}}$ columns and $N_{\text{row}}$ rows. Let $\overline{\boldsymbol d}^{\text{B}}_m\in\mathbb R^3$ and $\overline{\boldsymbol d}^{\text{R}}_n\in\mathbb R^3$ denote the unrotated local coordinates of the $m$-th BS transmit antenna and the $n$-th RIS reflecting element, respectively, and let $\boldsymbol d_0^{\text{B}}$ and $\boldsymbol d_0^{\text{R}}$ be their array centers in $\mathcal{G}$. Both BS and RIS employ a motorized mechanism to rotate the arrays as a whole. Based on the system's ISAC requirements, the proper rotation angles are determined by a central control center, which is connected to the motor controllers via fiber links\footnote{In practice, mechanical rotations can be updated on a slow timescale (e.g., sector-level adaptation or quasi-static geometry changes), whereas the BS digital precoder and RIS phases can be refreshed on the fast timescale of each channel-coherence block. Since the mechanical control involves only six scalar angles (3D Euler angles for the BS and RIS), the control signaling overhead is negligible compared with per-element electronic tuning. When the angles change, the array responses are updated deterministically; thus, standard pilot-based estimation and tracking remain applicable, and two-timescale designs using long-term statistical CSI for rotations are also feasible.}. 
Both the active array at BS and the passive RIS array rotate as a rigid body through Euler angles denoted by
\begin{align}
&\boldsymbol{r}^{\text{B}}\triangleq\big[r_{\mathrm{x}}^{\text{B}},r_{\mathrm{y}}^{\text{B}},r_{\mathrm{z}}^{\text{B}}\big]^{T}, \quad \boldsymbol{r}^{\text{R}}\triangleq\big[r_{\mathrm{x}}^{\text{R}},r_{\mathrm{y}}^{\text{R}},r_{\mathrm{z}}^{\text{R}}\big]^{T},
\end{align}
respectively, in their local CCS. In the following, we denote $\text{A}\in\{\text{B, R}\}$ as the notations for either BS or RIS array for brevity. Then, the corresponding rotation matrix generated by $\boldsymbol{r}^{\text{A}}$ represents the rotations about their local axis with the standard intrinsic $\mathrm{x}$-$\mathrm{y}$-$\mathrm{z}$ convention and is given by
\begin{align}
\label{eq:rotation_matrix_all}
\boldsymbol{R}\left( r_{\mathrm{x}}^{\text{A}},r_{\mathrm{y}}^{\text{A}},r_{\mathrm{z}}^{\text{A}} \right) =\boldsymbol{R}_{\mathrm{x}}\left( r_{\mathrm{x}}^{\text{A}} \right) \boldsymbol{R}_{\mathrm{y}}\left( r_{\mathrm{y}}^{\text{A}} \right) \boldsymbol{R}_{\mathrm{z}}\left( r_{\mathrm{z}}^{\text{A}} \right), 
\end{align}
where the rotation matrices on each axis are defined by
\begin{subequations} 
\label{eq:rotation_matrix}
\begin{align}
\label{eq:rotation_matrix_x}
&\boldsymbol{R}_{\mathrm{x}}\left( r_{\mathrm{x}}^{\text{A}} \right) =\left[ \begin{matrix}
	1&		0&		0\\
	0&		\cos r_{\mathrm{x}}^{\text{A}}&		-\sin r_{\mathrm{x}}^{\text{A}}\\
	0&		\sin r_{\mathrm{x}}^{\text{A}}&		\cos r_{\mathrm{x}}^{\text{A}}\\
\end{matrix} \right], 
\\
\label{eq:rotation_matrix_y}
&\boldsymbol{R}_{\mathrm{y}}\left( r_{\mathrm{y}}^{\text{A}} \right) =\left[ \begin{matrix}
	\cos r_{\mathrm{y}}^{\text{A}}&		0&		\sin r_{\mathrm{y}}^{\text{A}}\\
	0&		1&		0\\
	-\sin r_{\mathrm{y}}^{\text{A}}&		0&		\cos r_{\mathrm{y}}^{\text{A}}\\
\end{matrix} \right],
\\
\label{eq:rotation_matrix_z}
&\boldsymbol{R}_{\mathrm{z}}\left( r_{\mathrm{z}}^{\text{A}} \right) =\left[ \begin{matrix}
	\cos r_{\mathrm{z}}^{\text{A}}&		-\sin r_{\mathrm{z}}^{\text{A}}&		0\\
	\sin r_{\mathrm{z}}^{\text{A}}&		\cos r_{\mathrm{z}}^{\text{A}}&		0\\
	0&		0&		1\\
\end{matrix} \right].
\end{align}
\end{subequations}
The rotated antenna or element coordinates in $\mathcal{G}$ are then
\begin{subequations}
\label{eq:array_coordinate}
\begin{align}
\label{eq:BS_coordinate}
&\boldsymbol{d}_{m}^{\text{B}}\left( \boldsymbol{r}^{\text{B}} \right) =\boldsymbol{d}_{0}^{\text{B}}+\boldsymbol{R}\left( \boldsymbol{r}^{\text{B}} \right) \overline{\boldsymbol{d}}_{m}^{\text{B}},
\\
\label{eq:RIS_coordinate}
&\boldsymbol{d}_{n}^{\text{R}}\left( \boldsymbol{r}^{\text{R}} \right) =\boldsymbol{d}_{0}^{\text{R}}+\boldsymbol{R}\left( \boldsymbol{r}^{\text{R}} \right) \overline{\boldsymbol{d}}_{n}^{\text{R}}.
\end{align}
\end{subequations}
Note that rotations affect both (i) the position of antennas or elements in the phase term of the steering vector and (ii) the directivity gain of radiation patterns for the arrays, which are jointly captured through direction- and position-dependent gains introduced in the following channel modeling.

\subsection{Direction and Steering Vectors}
For a far-field direction parameterized by an elevation-azimuth pair $(\vartheta,\varphi)$ with elevation from the $xOy$ plane and azimuth from the $x$-axis, we define the departure direction vector as the unit propagation direction pointing from the array center to the far-field point
\begin{align}
\boldsymbol{u}(\vartheta,\varphi)
\triangleq[\cos\vartheta\cos\varphi,\cos\vartheta\sin\varphi,\sin\vartheta]^T.
\end{align} 
We adopt the geometry channel model to characterize the wireless propagation environment. Let $\mathcal{Q}_k=\{1,\dots,Q_k\}$, $\mathcal{L}_k=\{1,\dots,L_k\}$, and $\mathcal{P}=\{1,\dots,P\}$ denote the set of propagation paths between the $k$-th user and the BS, between the $k$-th user and the RIS, and between the RIS and the BS, respectively; specifically, we assume that the first paths in $\mathcal{Q}_k$, $\mathcal{L}_k$, and $\mathcal{P}$, corresponding to $l_k=1$, $q_k=1$ and $p=1$, respectively, are the line-of-sight (LoS) paths, while the remaining $L_k-1$, $Q_k-1$ and $P-1$ are the non-LoS paths. For the targeted sensing area with low-altitude UAVs, we consider merely the single LoS path with a potential reflected echo signal in each desired direction. 

Accordingly, we use the following indexed notational shorthands for distinct BS-user (BU), RIS-BS (RB), BS-RIS (BR), and RIS-user (RU) links, respectively, and define the direction vectors as follows
\begin{subequations}
\label{eq:direction_vector}
\begin{align}
&\boldsymbol{u}_{k,l}^{\text{BU}}\!=\!\left[ \cos( \vartheta _{k,l}^{\text{BU}} ) \cos( \varphi _{k,l}^{\text{BU}}) ,\cos( \vartheta _{k,l}^{\text{BU}} ) \sin( \varphi _{k,l}^{\text{BU}} ) ,\sin( \vartheta _{k,l}^{\text{BU}} ) \right] ^T \!,
\\
&\boldsymbol{u}_{p}^{\text{RB}}\!=\!\left[ \cos( \vartheta _{p}^{\text{RB}}) \cos ( \varphi _{p}^{\text{RB}}) ,\cos ( \vartheta _{p}^{\text{RB}}) \sin( \varphi _{p}^{\text{RB}}) ,\sin ( \vartheta _{p}^{\text{RB}}) \right] ^T \!,
\\
&\boldsymbol{u}_{p}^{\text{BR}}\!=\!\left[ \cos( \vartheta _{p}^{\text{BR}}) \cos( \varphi _{p}^{\text{BR}}) ,\cos( \vartheta _{p}^{\text{BR}}) \sin( \varphi _{p}^{\text{BR}} ) ,\sin ( \vartheta _{p}^{\text{BR}} ) \right] ^T \!,
\\
&\boldsymbol{u}_{k,q}^{\text{RU}}\!=\!\left[ \cos( \vartheta _{k,q}^{\text{RU}})\cos( \varphi _{k,q}^{\text{RU}}) ,\cos( \vartheta _{k,q}^{\text{RU}} ) \sin ( \varphi _{k,q}^{\text{RU}} ) ,\sin ( \vartheta _{k,q}^{\text{RU}} ) \right] ^T \!\!.
\end{align}
\end{subequations}
The associated steering vectors, jointly parameterized by the coordinates of the rotated elements in \eqref{eq:array_coordinate} and the direction vectors in \eqref{eq:direction_vector}, can be given by
\begin{subequations}
\label{eq:steering_vector_communication}
\begin{align}
&\!\boldsymbol{t}_{k,l}^{\text{BU}}( \boldsymbol{r}^{\text{B}} )\! =\!\big[ e^{j\frac{2\pi}{\lambda}( \boldsymbol{u}_{k,l}^{\text{BU}} ) ^T\boldsymbol{d}_{1}^{\text{B}}( \boldsymbol{r}^{\text{B}} )},\dots ,e^{j\frac{2\pi}{\lambda}( \boldsymbol{u}_{k,l}^{\text{BU}} ) ^T\boldsymbol{d}_{M}^{\text{B}}( \boldsymbol{r}^{\text{B}})} \big] ^T\!,\!\!
\\
&\!\boldsymbol{t}_{p}^{\text{BR}}( \boldsymbol{r}^{\text{B}}) \!=\!\big[ e^{j\frac{2\pi}{\lambda}( \boldsymbol{u}_{p}^{\text{BR}}) ^T\boldsymbol{d}_{1}^{\text{B}}( \boldsymbol{r}^{\text{B}})},\dots ,e^{j\frac{2\pi}{\lambda}( \boldsymbol{u}_{p}^{\text{BR}} ) ^T\boldsymbol{d}_{M}^{\text{B}}( \boldsymbol{r}^{\text{B}})} \big] ^T\!,\!\!
\\
&\!\boldsymbol{t}_{p}^{\text{RB}}( \boldsymbol{r}^{\text{R}} )\! =\!\big[ e^{j\frac{2\pi}{\lambda}( \boldsymbol{u}_{p}^{\text{RB}} ) ^T\boldsymbol{d}_{1}^{\text{R}}( \boldsymbol{r}^{\text{R}})},\dots ,e^{j\frac{2\pi}{\lambda}( \boldsymbol{u}_{p}^{\text{RB}} ) ^T\boldsymbol{d}_{N}^{\text{R}}( \boldsymbol{r}^{\text{R}})} \big] ^T\!,\!\!
\\
&\!\boldsymbol{t}_{k,q}^{\text{RU}}( \boldsymbol{r}^{\text{R}})\! =\!\big[ e^{j\frac{2\pi}{\lambda}( \boldsymbol{u}_{k,q}^{\text{RU}} ) ^T\boldsymbol{d}_{1}^{\text{R}}( \boldsymbol{r}^{\text{R}} )},\dots ,e^{j\frac{2\pi}{\lambda}( \boldsymbol{u}_{k,q}^{\text{RU}}) ^T\boldsymbol{d}_{N}^{\text{R}}( \boldsymbol{r}^{\text{R}})} \big] ^T\!.\!\!
\end{align}
\end{subequations}
Similarly, for a target direction $(\vartheta_a,\varphi_a)$ within the target sensing area $\mathcal{A}$, the direction steering vectors for BS-target (BT) and RIS-target (RT) links can be respectively given by
\begin{subequations}
\begin{align}
&\!\boldsymbol{u}_a^{\text{BT}}\!\!=\!\!\left[ \cos(\vartheta_{a}^{\text{BT}})\! \cos( \varphi_{a}^{\text{BT}}), \cos(\vartheta_{a}^{\text{BT}})\! \sin(\varphi_{a}^{\text{BT}}), \sin(\vartheta_{a}^{\text{BT}}) \right] ^T\!\!,\!\!
\\
&\!\boldsymbol{u}_a^{\text{RT}}\!\!=\!\!\left[ \cos(\vartheta_{a}^{\text{RT}})\! \cos( \varphi_{a}^{\text{RT}}), \cos(\vartheta_{a}^{\text{RT}})\! \sin(\varphi_{a}^{\text{RT}}) ,\sin(\vartheta_{a}^{\text{RT}}) \right] ^T\!\!,\!\!
\end{align}
\end{subequations}
and
\begin{subequations}
\label{eq:steering_vector_sensing}
\begin{align}
&\boldsymbol{t}_a^{\text{BT}}(\boldsymbol{r}^{\text{B}}) =\big[ e^{j\frac{2\pi}{\lambda}( \boldsymbol{u}_{a}^{\text{BT}}) ^T\boldsymbol{d}_{1}^{\text{B}}( \boldsymbol{r}^{\text{B}})},\dots ,e^{j\frac{2\pi}{\lambda}( \boldsymbol{u}_{a}^{\text{BT}}) ^T\boldsymbol{d}_{M}^{\text{B}}( \boldsymbol{r}^{\text{B}})} \big] ^T,
\\
&\boldsymbol{t}_a^{\text{RT}}(\boldsymbol{r}^{\text{R}}) =\big[ e^{j\frac{2\pi}{\lambda}(\boldsymbol{u}_{a}^{\text{RT}}) ^T\boldsymbol{d}_{1}^{\text{R}}( \boldsymbol{r}^{\text{R}} )},\dots ,e^{j\frac{2\pi}{\lambda}( \boldsymbol{u}_{a}^{\text{RT}}) ^T\boldsymbol{d}_{N}^{\text{R}}( \boldsymbol{r}^{\text{R}})} \big] ^T.
\end{align}
\end{subequations}
As shown in \eqref{eq:steering_vector_communication} and \eqref{eq:steering_vector_sensing}, array rotations, which generate position-dependent effects on the steering vector's phase term, can be exploited to reshape channel responses.

\subsection{Rotation-Aware Multipath Channels}
To capture rotation-dependent directivity in the radiation pattern of BS and RIS arrays, we include non-negative gain terms $G^{\text{B}}\left( \boldsymbol{r}^{\text{B}} \right)$ and $G^{\text{R}}\left( \boldsymbol{r}^{\text{R}} \right)$, which model directional antenna and element patterns under rotation. Let $\bar{\boldsymbol{n}}=[0,0,1]^T$ and 
\begin{align}
\boldsymbol{n}(\boldsymbol{r}^{\text{A}})=\boldsymbol{R}(\boldsymbol{r}^{\text{A}})\bar{\boldsymbol{n}},
\end{align}
be the unrotated and rotated array boresight, respectively, we adopt a practical directivity model
\begin{subequations}
\label{eq:directivity_gain}
\begin{align}
&G^{\text{B}}\left( \boldsymbol{r}^{\text{B}};\boldsymbol{u}^{\text{B}} \right) \!=\!\begin{cases}
	G_{0}^{\text{B}}\left( \boldsymbol{n}( \boldsymbol{r}^{\text{B}}) ^T\boldsymbol{u}^{\text{B}}\right) ^{b^{\text{B}}}, \boldsymbol{n}( \boldsymbol{r}^{\text{B}} ) ^T\boldsymbol{u}^{\text{B}}>0,\!\!\!\!\!\\
	0,         \qquad \qquad \qquad \quad \,\, \text{otherwise}.\!\!\!\!\!\\
\end{cases} \!\!\!
\\
&G^{\text{R}}\left( \boldsymbol{r}^{\text{R}};\boldsymbol{u}^{\text{R}} \right) \!=\!\begin{cases}
	G_{0}^{\text{R}}\left( \boldsymbol{n}( \boldsymbol{r}^{\text{R}} ) ^T\boldsymbol{u}^{\text{R}} \right) ^{b^{\text{R}}}, \boldsymbol{n} ( \boldsymbol{r}^{\text{R}}) ^T\boldsymbol{u}^{\text{R}}>0\!\!\!\!\!\\
	0,      \qquad \qquad \qquad \quad \,\,  \text{otherwise},\!\!\!\!\!\\
\end{cases} \!\!\!
\end{align}
\end{subequations}
where $G_{0}^{\text{B}}$ and $G_{0}^{\text{R}}$ denotes the maximum gain achieved when the signal aligns with the boresight, and $b^{\text{B}}$ and $b^{\text{R}}$ are the directivity factors controlling the main-lobe width, following from directivity normalization over the full sphere. Then, with given numbers of resolvable paths $L_k$, $P$, and $Q_k$, and small-scale complex gains $\delta_{k,l}$, $\gamma_p$, and $\psi_{k,q}$ that absorb path loss and fading, the BS-to-user channel $\boldsymbol{h}_k\left( \boldsymbol{r}^{\text{B}} \right) \in \mathbb{C}^M$, BS-to-RIS channel $\boldsymbol{B}\left( \boldsymbol{r}^{\text{B}},\boldsymbol{r}^{\text{R}} \right) \in \mathbb{C}^{M\times N}$, and RIS-to-user channel $\boldsymbol{g}_k\left( \boldsymbol{r}^{\text{R}} \right) \in \mathbb{C}^N$ are respectively expressed as
\begin{subequations}
\begin{align}
&\boldsymbol{h}_k\left( \boldsymbol{r}^{\text{B}} \right) =\sum_{l=1}^{L_k}{\delta_{k,l}\sqrt{G_{k,l}^{\text{B}}\left( \boldsymbol{r}^{\text{B}} \right)}}\boldsymbol{t}_{k,l}^{\text{BU}}\left( \boldsymbol{r}^{\text{B}} \right), 
\\
&\boldsymbol{B}\left( \boldsymbol{r}^{\text{B}},\boldsymbol{r}^{\text{R}} \right) =\sum_{p=1}^P{\gamma_p\sqrt{G_{p}^{\text{B}}\left( \boldsymbol{r}^{\text{B}},\boldsymbol{r}^{\text{R}} \right) G_{p}^{\text{R}}\left( \boldsymbol{r}^{\text{B}},\boldsymbol{r}^{\text{R}} \right)}} \nonumber
\\
&\qquad\qquad \qquad \quad\,\,\times \boldsymbol{t}_{p}^{\text{BR}}\left( \boldsymbol{r}^{\text{B}} \right) \left( \boldsymbol{t}_{p}^{\text{RB}}\left( \boldsymbol{r}^{\text{R}} \right) \right) ^H,
\\
&\boldsymbol{g}_k\left( \boldsymbol{r}^{\text{R}} \right) =\sum_{q=1}^{Q_k}{\psi_{k,q}\sqrt{G_{k,q}^{\text{R}}\left( \boldsymbol{r}^{\text{R}} \right)}}\boldsymbol{t}_{k,q}^{\text{RU}}\left( \boldsymbol{r}^{\text{R}} \right). 
\end{align}
\end{subequations}
Taking into account direct and cascaded RIS-enabled links, the effective composite channel from BS to the $k$-th user is
\begin{align}
\label{eq:channel_BS_user}
\boldsymbol{f}_k ( \boldsymbol{r}^{\text{B}}\!,\boldsymbol{r}^{\text{R}}\!,\boldsymbol{\theta } ) \!=\!\boldsymbol{h}_k( \boldsymbol{r}^{\text{B}} )\! +\!\boldsymbol{B}( \boldsymbol{r}^{\text{B}},\boldsymbol{r}^{\text{R}} ) \mathrm{diag}( \boldsymbol{\theta } ) \boldsymbol{g}_k( \boldsymbol{r}^{\text{R}} ), \!\!
\end{align}
where $\boldsymbol\theta=[\theta_1,\ldots,\theta_N]^T$ collects the RIS reflecting phase shift coefficients, typically being passive with $|\theta_n|=1$. Similarly, for sensing, define the effective steering toward a target direction $(\vartheta_a,\varphi_a)$ as
\begin{align}
\label{eq:channel_BS_target}
\boldsymbol{f}_{\text{S},a}( \boldsymbol{r}^{\text{B}},\boldsymbol{r}^{\text{R}},\boldsymbol{\theta }) =&\sqrt{G_{a}^{\text{B}}\left( \boldsymbol{r}^{\text{B}} \right)} \boldsymbol{t}_{a}^{\text{BT}}\left( \boldsymbol{r}^{\text{B}} \right) \nonumber
+\!\boldsymbol{B}( \boldsymbol{r}^{\text{B}},\boldsymbol{r}^{\text{R}} ) \mathrm{diag}( \boldsymbol{\theta } ) 
\\
&\times \sqrt{G_{a}^{\text{R}}\left( \boldsymbol{r}^{\text{R}} \right)} \boldsymbol{t}_{a}^{\text{RT}}( \boldsymbol{r}^{\text{R}} ). 
\end{align}

\subsection{Transmit Signal, Communication SINR, and Sensing NMSE}
The BS transmits an ISAC signal
\begin{align}
\boldsymbol{x}=\boldsymbol{W}_{\text{C}}\boldsymbol{s}_{\text{C}}+\boldsymbol{W}_{\text{S}}\boldsymbol{s}_{\text{S}},
\end{align}
where $\boldsymbol{W}_{\text{C}}=[\boldsymbol{w}_1,\dots,\boldsymbol{w}_K\boldsymbol] \in \mathbb{C}^{M\times K}$ collects $K$ communication beams, $\boldsymbol{W}_{\text{S}}\in \mathbb{C}^{M\times M}$ contains dedicated sensing beams, $\boldsymbol{s}_\text{C}=[s_{\text{C},1},\dots,s_{\text{C},K}]^T$ denotes the communication symbols, and $\boldsymbol{s}_\text{S}$ denotes the sensing signals, with $\mathbb{E}\left[ \boldsymbol{s}_{\text{S}}\boldsymbol{s}_{\text{S}}^{H} \right] =\boldsymbol{I}_M$, $\boldsymbol{s}_{\text{C}}\sim \mathcal{C}\mathcal{N}\left( \boldsymbol{0},\boldsymbol{I}_K \right)$, and $\mathbb{E}\left[ \boldsymbol{s}_{\text{S}}\boldsymbol{s}_{\text{C}}^{H} \right] =0$.
Let $\boldsymbol{W}\triangleq \left[ \boldsymbol{W}_{\text{C}},\boldsymbol{W}_{\text{S}} \right]$ denotes the BS ISAC transmit beamforming matrix, $\boldsymbol{s}\triangleq \left[ \boldsymbol{s}_{\text{C}}^{T},\boldsymbol{s}_{\text{S}}^{T} \right]^T$, the received signal at the communication user $k$ is written as
\begin{subequations}
\begin{align}
c_k&=\boldsymbol{f}_{k}^{H}( \boldsymbol{r}^{\text{B}},\boldsymbol{r}^{\text{R}},\boldsymbol{\theta } ) \boldsymbol{Ws}
\\
&=\boldsymbol{f}_{k}^{H}( \boldsymbol{r}^{\text{B}},\boldsymbol{r}^{\text{R}},\boldsymbol{\theta } ) \boldsymbol{w}_{k}s_{\text{C},k}+\sum_{i=1,i\ne k}^{K}{\boldsymbol{f}_{k}^{H}( \boldsymbol{r}^{\text{B}},\boldsymbol{r}^{\text{R}},\boldsymbol{\theta } ) \boldsymbol{w}_{i}^{}}s_{\text{C},i} \nonumber
\\
&\quad +\boldsymbol{f}_{k}^{H}( \boldsymbol{r}^{\text{B}},\boldsymbol{r}^{\text{R}},\boldsymbol{\theta } ) \boldsymbol{W}_{\text{S}}\boldsymbol{s}_{\text{S}} +n_k,
\end{align}
\end{subequations}
where $n_k\sim \mathcal{C}\mathcal{N}\left( 0,\sigma _{k}^{2} \right) $ is the complex additive white Gaussian noise. Taking further $\boldsymbol{W}\triangleq \left[ \boldsymbol{w}_1,\dots ,\boldsymbol{w}_{K+M} \right]$, where $\boldsymbol{w}_i$ is the $i$-th column of $\boldsymbol{W}$, the SINR of the $k$-th user is
\begin{align}
\text{SINR}_k=\frac{\big| \boldsymbol{f}_{k}^{H}\left( \boldsymbol{r}^{\text{B}},\boldsymbol{r}^{\text{R}},\boldsymbol{\theta } \right) \boldsymbol{w}_{k}^{} \big|^2}{\sum_{i=1,i\ne k}^{K+M}{\big| \boldsymbol{f}_{k}^{H}\left( \boldsymbol{r}^{\text{B}},\boldsymbol{r}^{\text{R}},\boldsymbol{\theta } \right) \boldsymbol{w}_{i} \big|^2}+\sigma _{k}^{2}},
\end{align}

For sensing, the transmit beampattern toward $(\vartheta_a,\varphi_a)$ is
\begin{subequations}
\label{eq:transmit_beam_pattern}
\begin{align}
\mathcal{P}( \vartheta_a ,\varphi_a ) &=\mathbb{E}( | \boldsymbol{f}_{\text{S},a}^{H}\left( \boldsymbol{r}^{\text{B}},\boldsymbol{r}^{\text{R}},\boldsymbol{\theta } \right) \boldsymbol{Ws} |^2 ) 
\\
&=\boldsymbol{f}_{\text{S},a}^{H}( \boldsymbol{r}^{\text{B}},\boldsymbol{r}^{\text{R}},\boldsymbol{\theta } ) \boldsymbol{WW}^H\boldsymbol{f}_{\text{S},a}( \boldsymbol{r}^{\text{B}},\boldsymbol{r}^{\text{R}},\boldsymbol{\theta } ). 
\end{align}
\end{subequations}
Using \eqref{eq:channel_BS_user}, \eqref{eq:transmit_beam_pattern} expands to
\begin{align}
&\mathcal{P}( \vartheta _a,\varphi _a ) \nonumber
\\
&=( \boldsymbol{t}_{a}^{\text{BT}}( \boldsymbol{r}^{\text{B}})) ^H\boldsymbol{WW}^H\boldsymbol{t}_{a}^{\text{BT}}( \boldsymbol{r}^{\text{B}} ) 
+\!2\sqrt{G_{a}^{\text{B}}( \boldsymbol{r}^{\text{B}}) G_{a}^{\text{R}}( \boldsymbol{r}^{\text{R}})}
\nonumber
\\
&\quad \quad\times\text{Re}\{ ( \boldsymbol{t}_{a}^{\text{BT}}( \boldsymbol{r}^{\text{B}} ) ) ^H\boldsymbol{WW}^H\boldsymbol{B}( \boldsymbol{r}^{\text{B}},\boldsymbol{r}^{\text{R}}) \mathrm{diag}( \boldsymbol{\theta }) \boldsymbol{t}_{a}^{\text{RT}}( \boldsymbol{r}^{\text{R}}) \} \nonumber
\\
&\quad             +\sqrt{G_{a}^{\text{B}}( \boldsymbol{r}^{\text{B}}) G_{a}^{\text{R}}( \boldsymbol{r}^{\text{R}} )}( \boldsymbol{B}\left( \boldsymbol{r}^{\text{B}},\boldsymbol{r}^{\text{R}} \right) \mathrm{diag}( \boldsymbol{\theta } ) \boldsymbol{t}_{a}^{\text{RT}}( \boldsymbol{r}^{\text{R}} ) ) ^H \nonumber
\\
& \quad \quad \times \boldsymbol{WW}^H\boldsymbol{B}( \boldsymbol{r}^{\text{B}},\boldsymbol{r}^{\text{R}}) \mathrm{diag}( \boldsymbol{\theta } ) \boldsymbol{t}_{a}^{\text{RT}}( \boldsymbol{r}^{\text{R}} ) .
\end{align}

Given a discretized grid $\{(\vartheta_a,\varphi_a)\}_{a=1}^A,a\in\{1,\dots,A\}$ with $A$ sampled points in total, which samples the azimuth and elevation angular domain over the target sensing area $\mathcal{A}$, along with the desired beampattern $\mathcal{P}_{\text{d}}(\vartheta_a,\varphi_a)$, the mean squared error (MSE) between $\mathcal{P}$ and $\mathcal{P}_\text{d}$ to evaluate the similarity of the sensing beampattern is defined as \cite{ref31}
\begin{align}
\text{MSE}\!=\!\frac{1}{A}\!\sum_{a=1}^A\!{\left| \mathcal{P}( \vartheta _a,\varphi _a;\boldsymbol{r}^{\text{B}},\boldsymbol{r}^{\text{R}},\boldsymbol{\theta },\boldsymbol{W})\! -\!\mathcal{P}_{\text{d}}( \vartheta_a,\varphi_a ) \right|^2}\!.\!\!\!
\end{align}
A typical spotlight-type desired beampattern is given by
\begin{align}
\mathcal{P}_{\text{d}}\left( \vartheta _a,\varphi _a \right) =\begin{cases}
	1, a\in \mathcal{F}\\
	0, \text{otherwise}\\
\end{cases}
\end{align}
where $\mathcal{F}$ is the set of focused grid points in the region $\mathcal{A}$. 
While MSE is intuitive, it is scale-sensitive: the same beampattern shape may lead to very different MSE values when the overall beampattern level fluctuates due to transmit-power changes, propagation loss, blockage, or orientation-dependent element gains, which are prevalent in low-altitude geometries. To obtain a dimensionless and scale-robust metric that emphasizes shape fidelity rather than absolute amplitude, we adopt the NMSE metric with a scaling parameter $\iota>0$ as
\begin{align}
\label{eq:nmse_def}
\text{NMSE}\triangleq\frac{\sum_{a=1}^A{\left| \mathcal{P}_a-\iota \mathcal{P}_{d,a} \right|^2}}{\sum_{a=1}^A{\left| \iota \mathcal{P}_{d,a} \right|^2}}=
\frac{\left\|\boldsymbol{p}-\iota \boldsymbol{p}_{\text{d}}\right\|_2^2}{\left\|\iota \boldsymbol{p}_{\text{d}}\right\|_2^2}.
\end{align}
where $\boldsymbol{p}=[\mathcal{P}_1,\ldots,\mathcal{P}_A]^{T}$ and $\boldsymbol{p}_{\text{d}}=[\mathcal{P}_{\text{d},1},\ldots,\mathcal{P}_{\text{d},A}]^{T}$. 
The scaling parameter $\iota$ decouples the global amplitude mismatch from the spatial shape mismatch: even when $\mathcal{P}_a$ is uniformly scaled by channel loss or array gain variations, $\text{NMSE}(\iota)$ can remain small as long as the angular shape matches. Moreover, using $\|\iota \boldsymbol{p}_{\text{d}}\|_2^2$ in the denominator avoids the trivial solution $\iota=0$, which would artificially reduce the numerator, thereby enforcing a meaningful normalization\footnote{The adopted beampattern matching NMSE is a design-friendly proxy for sensing quality: a smaller NMSE indicates better mainlobe fidelity in the intended angular sectors and reduced leakage or sidelobes elsewhere, which typically improves the received peak-to-maximum-sidelobe ratio (PSLR) and signal-to-clutter-plus-noise ratio (SCNR), and the reliability of detection or estimation in the region of interest}.

For fixed $(\boldsymbol W,\boldsymbol\theta,\boldsymbol r^{\text{B}},\boldsymbol r^{\text{R}})$, we update $\iota$ by minimizing $\text{NMSE}(\iota)$ in closed form. Let $c=\|\boldsymbol{p}\|_2^2$, $d=\boldsymbol{p}_{\text{d}}^{T}\boldsymbol{p}$, and $e=\|\boldsymbol{p}_{\text{d}}\|_2^2$. When $d>0$, the minimizer is
\begin{equation}
\iota^\star=\arg\min_{\iota>0}\text{NMSE}(\iota)=\frac{\|\boldsymbol{p}\|_2^2}{\boldsymbol{p}_{\text{d}}^{T}\boldsymbol{p}}.
\label{eq:iota_star}
\end{equation}
In practice, to avoid numerical issues when $\boldsymbol{p}_{\text{d}}^{T}\boldsymbol{p}$ is very small (e.g., at initialization), we use a safeguarded update $\iota^\star=\|\boldsymbol{p}\|_2^2/(\boldsymbol{p}_{\text{d}}^{T}\boldsymbol{p}+\varepsilon)$ with a small $\varepsilon>0$. With $\iota=\iota^\star$, the NMSE in \eqref{eq:nmse_def} measures how well the synthesized pattern matches the desired pattern up to an overall scaling, which is particularly suitable when (i) the absolute beampattern level is affected by the BS power budget and propagation loss, (ii) the sensing objective primarily concerns mainlobe or sidelobe shape rather than absolute amplitude, and (iii) the system expects the sensing metric to be comparable across different system settings, e.g., $P_{\text{B}}$, $M$, $N$, and user or target geometries.

\vspace{-6pt}
\subsection{Problem Formulation }
Accordingly, we jointly optimize the BS and RIS rotations, the RIS phase shifts, and the BS transmit beamforming to maximize a weighted-sum utility of downlink communication sum rate and sensing beampattern fidelity. Specifically, we consider
\begin{subequations}
\begin{align}
\text{(P1)}:\underset{\boldsymbol{r}^{\text{B}},\boldsymbol{r}^{\text{R}},\boldsymbol{\theta },\boldsymbol{W}}{\max}&\sum_{k=1}^K{\log _2\left( 1+\text{SINR}_k \right)} -\rho \text{NMSE}
\label{p1:objective}
\\
\text{s.t.} \quad &\|\boldsymbol{W}\|_{F}^{2}=\mathrm{Tr}(\boldsymbol{WW}^H) \le P_{\text{B}},
\label{p1:power}
\\
&|\theta _n|=1,\forall n\in \mathcal{N},
\label{p1:phase}
\\
&r _{\mathrm{i}}^{\text{B}}\!\in \left[ r _{\mathrm{i},\min}^{\text{B}},r _{\mathrm{i},\max}^{\text{B}} \right] ,\mathrm{i}\in \{\mathrm{x},\mathrm{y},\mathrm{z}\},
\label{p1:Bbox}
\\
&r _{\mathrm{i}}^{\text{R}}\!\in \left[ r _{\mathrm{i},\min}^{\text{R}},r _{\mathrm{i},\max}^{\text{R}} \right] ,\mathrm{i}\in \{\mathrm{x},\mathrm{y},\mathrm{z}\},
\label{p1:Rbox}
\end{align}
\end{subequations}
where $\rho\ge 0$ in \eqref{p1:objective} is a trade-off weight controlling the relative emphasis on communication throughput versus sensing beampattern matching, $P_{\text{B}}$ in \eqref{p1:power} is the BS power budget, the unit-modulus constraint \eqref{p1:phase} enforces a passive RIS, and the box constraints \eqref{p1:Bbox} and \eqref{p1:Rbox} capture mechanical limits of the array rotations at BS and RIS, respectively. Note that the proposed formulation in \eqref{p1:objective} directly targets the fundamental ISAC trade-off by optimizing a scalarized Pareto objective.

Compared with a strict threshold form of constraint on $\text{NMSE}$, the weighted-sum design offers three practical advantages. First, in low-altitude scenarios, the sensing pattern requirement may become temporarily difficult or even infeasible under fast geometry variations and strong multiuser interference, while the weighted-sum objective avoids infeasibility, being robust to geometry variation, and yields a graceful performance trade-off. Second, hard-threshold formulations typically require penalty or dual updates and may introduce non-smooth hinge-type terms, which can complicate algorithm design and slow convergence, while a weighted sum leads to a single-level smooth objective, enabling simpler tuning and more stable optimization. Third, by sweeping $\rho$, one can efficiently explore the Pareto frontier between the sum rate and NMSE, which provides a comprehensive characterization of the communication-sensing trade-off beyond enforcing a single scenario-dependent threshold.

From the formulated problem, we observe that the rotation variables $(\boldsymbol r^{\text B},\boldsymbol r^{\text R})$ provide a \textit{geometric} control that affects the steering vectors and directive gain, thereby complementing the \textit{electromagnetic} control offered by active and passive beamforming at the RIS and BS arrays; consequently, jointly optimizing these blocks yields a significant DoF for low-altitude ISAC. However, (P1) is highly non-convex and challenging to solve directly due to unit-modulus constraints, rotation-induced nonlinearities in the steering vector and directive gain, and the coupling among $\boldsymbol{r}^{\text{B}}$, $\boldsymbol{r}^{\text{R}}$, $\boldsymbol{\theta}$, and $\boldsymbol{W}$ in both SINR and MSE expressions.

\section{Proposed Algorithm}

In this section, we develop an efficient algorithm to tackle the non-concave objective and the non-convex constraints in (P1), where the optimization variables are tightly coupled through both the multiuser SINRs and the sensing NMSE.

\vspace{-6pt}
\subsection{AO framework}

A key observation is that (P1) becomes considerably more tractable when optimizing one variable block at a time. Hence, we resort to an AO framework that alternately updates three blocks in an iterative manner: the BS transmit beamforming matrix $\boldsymbol W$, the RIS passive beamforming vector $\boldsymbol\theta$, and the array rotations $(\boldsymbol r^{\text{B}},\boldsymbol r^{\text{R}})$. 

Importantly, the sensing metric is $\text{NMSE}(\iota)$ in \eqref{eq:nmse_def}, where the scaling parameter $\iota$ is introduced to decouple global amplitude mismatch from angular shape mismatch. We update $\iota$ once per AO outer iteration using the closed-form minimizer $\iota^\star$ in \eqref{eq:iota_star}, and keep it fixed when solving the three inner subproblems for $\boldsymbol W$, $\boldsymbol\theta$, and $(\boldsymbol r^{\text{B}},\boldsymbol r^{\text{R}})$. This outer update is computationally negligible and does not introduce additional constraints. Moreover, since $\iota^\star$ minimizes $\text{NMSE}(\iota)$ for fixed $(\boldsymbol W, \boldsymbol\theta, \boldsymbol r^{\text{B}},\boldsymbol r^{\text{R}})$, and the sum-rate term is independent of $\iota$, updating $\iota$ with $\iota^\star$ makes the objective value in \eqref{p1:objective} non-decreasing. Therefore, the $\iota$-update does not undermine the monotonic improvement of the overall AO procedure; it simply refines the sensing metric by removing the best-fit scaling mismatch at the current iterate.

Then, in each iteration, the resulting three subproblems are handled by: i) a QT combined with a MM step for updating $\boldsymbol W$, ii) a Riemannian optimization method over the complex circle manifold for updating $\boldsymbol\theta$, and iii) a projected gradient-based method for updating $(\boldsymbol r^{\text{B}},\boldsymbol r^{\text{R}})$ under box constraints, respectively. Due to the intrinsic non-convexity of (P1), global optimality is generally intractable. The proposed AO-based algorithm aims at computing a locally optimal or suboptimal solution that achieves a favorable communication-sensing trade-off. In practice, such a suboptimal solution is meaningful because it is obtained with low complexity, respects all hardware constraints such as unit-modulus phases and mechanical rotation limits, and can be efficiently warm-started and re-optimized as the low-altitude geometry evolves.

\subsection{QT- and MM-based Transmit Beamforming}
With $(\boldsymbol{\theta},\boldsymbol r^{\text{B}},\boldsymbol r^{\text{R}})$ and $\iota$ fixed, the subproblem for optimizing $\boldsymbol W$ becomes
\begin{align}
\text{(P2)}: \underset{\boldsymbol W}{\max}\,
&\sum_{k=1}^{K}\log_2\left(1+\mathrm{SINR}_k(\boldsymbol W)\right)
-\rho\text{NMSE}(\boldsymbol W)
\label{p2:W_nmse_obj}
\\
\text{s.t.} \quad & \eqref{p1:power} , \nonumber
\end{align}
and the SINR and NMSE expressions reduce to 
\begin{subequations}
\begin{align}
\text{SINR}_k(\boldsymbol{W}) &=\frac{|\boldsymbol{f}_{k}^{H}\boldsymbol{w}_{k}^{}|^2}{\sum_{i=1,i\ne k}^{K+M}{|\boldsymbol{f}_{k}^{H}\boldsymbol{w}_{i}^{}|^2}+\sigma _{k}^{2}},
\label{eq:sinr_W}
\\
\text{NMSE}(\boldsymbol W)&=
\frac{\sum_{a=1}^{A}\big| \|\boldsymbol W^{H}\boldsymbol f_{\text{S},a}\|_2^2-\iota \mathcal{P}_{d,a} \big|^2}
{\sum_{a=1}^{A}\left|\iota \mathcal{P}_{d,a}\right|^2} 
\\
&= \frac{1}{D}\sum_{a=1}^{A}\big| \|\boldsymbol W^{H}\boldsymbol f_{\text{S},a}\|_2^2- \bar{\mathcal{P}}_{d,a} \big|^2,
\label{eq:nmse_W}
\end{align}
\end{subequations}
where we denote $D\triangleq\sum_{a=1}^{A}\left|\iota \mathcal{P}_{d,a}\right|^2$ and $\bar{\mathcal{P}}_{d,a}\triangleq \iota \mathcal{P}_{d,a}$ determined by the fixed $\iota$ for brevity. Here, $\boldsymbol f_k$ denotes the effective communication channel of user $k$, and $\boldsymbol f_{\text{S},a}$ denotes the effective sensing steering vector at grid point $a$, both determined by the fixed $(\boldsymbol\theta,\boldsymbol r^{\text{B}},\boldsymbol r^{\text{R}})$. As the power constraint \eqref{p1:power} is convex, the non-concavity of \eqref{p2:W_nmse_obj} mainly stems from the fractional SINR terms in \eqref{eq:sinr_W} and the fourth-order dependence of the sensing loss $\|\boldsymbol W^H\boldsymbol f_{\text{S},a}\|_2^2$ in \eqref{eq:nmse_W}. In the sequel, we first apply the QT to reformulate the sum-rate into a tractable form, and then employ an MM step to construct a tight surrogate for the sensing-related quartic term, leading to a closed-form update of $\boldsymbol W$ with bisection search.

\subsubsection{Sensing MSE term with MM-based second-order upper bound}
For brevity, let 
\begin{subequations}
\label{W_sensing_MSE}
\begin{align}
g_a(\boldsymbol{W}) 
&\triangleq \big( \lVert \boldsymbol{W}^H\boldsymbol{f}_{\text{S},a} \rVert _{2}^{2}-\bar{\mathcal{P}}_{\text{d},a} \big) ^2,
\\
&=\big( \text{Tr} ( \boldsymbol{W}^H\boldsymbol{S}_a\boldsymbol{W}) -\bar{\mathcal{P}}_{\text{d},a} \big) ^2,
\end{align}
\end{subequations}
with the rank-one positive semi-definite matrix defined by
\begin{align}
\label{eq:S_a}
\boldsymbol{S}_a\triangleq\boldsymbol{f}_{\text{S},a}\boldsymbol{f}_{\text{S},a}^H\succeq\boldsymbol{0}, 
\end{align}
we have the gradient of \eqref{W_sensing_MSE} being
\begin{subequations}
\begin{align}
\label{eq:G_a}
\boldsymbol{G}_a\left( \boldsymbol{W} \right) 
&\triangleq \nabla g_a\left( \boldsymbol{W} \right) 
\\
&=4\big( \mathrm{Tr}( \boldsymbol{W}^H\boldsymbol{S}_a\boldsymbol{W} ) -\bar{\mathcal{P}}_{\text{d},a} \big) \boldsymbol{S}_a\boldsymbol{W}.
\end{align}
\end{subequations}
At the current iterate $\boldsymbol{W}^{(t)}$, we apply the second-order Taylor expansion (also known as the Descent Lemma \cite{ref26}) to construct a convex upper-bound surrogate function for $g_a(\boldsymbol{W})$ as
\begin{subequations}
\label{eq:W_sensing_bound}
\begin{align}
g_a\left( \boldsymbol{W} \right) &\le g_a\big( \boldsymbol{W}^{\left( t \right)} \big) +\text{Re}\big\{ \text{Tr}\big( ( \boldsymbol{W}-\boldsymbol{W}^{(t)}) ^H\boldsymbol{G}_a (\boldsymbol{W}^{(t)}) \big) \big\}  \nonumber
\\
& \quad +\frac{L_{a}^{\text{lip}}}{2}\big\| \boldsymbol{W}-\boldsymbol{W}^{(t)} \big\| _{F}^{2}
\\
& \triangleq \mathcal{U}_a\big( \boldsymbol{W}\mid \boldsymbol{W}^{(t)} \big) ,
\end{align}
\end{subequations}
where $L_a^{\text{lip}}$ is any Lipschitz constant on the Frobenius ball
$\{\boldsymbol{W}:\|\boldsymbol{W}\|_F^2\le P_{\text{B}}\}$ for $\nabla g_a( \boldsymbol{W})$. Using matrix-norm inequalities, one obtains a choice for a global Lipschitz constant
\begin{align}
\label{eq:L_a}
L_{a}^{\text{lip}}=12P_{\text{B}}\lVert \boldsymbol{S}_a \rVert _{2}^{2}+4\bar{\mathcal{P}}_{\text{d},a}\lVert \boldsymbol{S}_a \rVert _{2}^{},
\end{align}
with $\|\boldsymbol{S}_a\|_2=\|\boldsymbol{f}_{\text{S},a}\|_2^2$.

\begin{IEEEproof} 
For any two points $\boldsymbol{W}_1$ and $\boldsymbol{W}_2$ on the function $g_a( \boldsymbol{W})$, according to matrix norm inequalities, we have
\begin{align} \label{eq:proof_norm_inequality}
&\lVert \nabla g_a\left( \boldsymbol{W}_1 \right) -\nabla g_a\left( \boldsymbol{W}_2 \right) \rVert \nonumber
\\
&=4\big\|( \mathrm{Tr}( \boldsymbol{W}_{1}^{H}\boldsymbol{S}_a\boldsymbol{W}_1 ) -\bar{\mathcal{P}}_{\text{d},a} ) \boldsymbol{S}_a( \boldsymbol{W}_1-\boldsymbol{W}_2 ) \nonumber
\\
&\quad +( \mathrm{Tr}( \boldsymbol{W}_{1}^{H}\boldsymbol{S}_a\boldsymbol{W}_1 ) -\mathrm{Tr}( \boldsymbol{W}_{2}^{H}\boldsymbol{S}_a\boldsymbol{W}_2 ) ) \boldsymbol{S}_a\boldsymbol{W}_2\big\| \nonumber
\\
&\le 4\lVert \boldsymbol{S}_a \rVert _2  \big( |\mathrm{Tr}( \boldsymbol{W}_{1}^{H}\boldsymbol{S}_a\boldsymbol{W}_1 ) -\bar{\mathcal{P}}_{\text{d},a}|\lVert \boldsymbol{W}_1-\boldsymbol{W}_2 \rVert _F \nonumber
\\
&\quad  +| \mathrm{Tr}( \boldsymbol{W}_{1}^{H}\boldsymbol{S}_a\boldsymbol{W}_1 ) \!-\!\mathrm{Tr}( \boldsymbol{W}_{2}^{H}\boldsymbol{S}_a\boldsymbol{W}_2 ) |\lVert \boldsymbol{W}_2 \rVert _F\big). \!
\end{align}
Then, we construct the following inequalities for terms related to traces in \eqref{eq:proof_norm_inequality}
\begin{align} 
&| \mathrm{Tr}( \boldsymbol{W}_{1}^{H}\boldsymbol{S}_a\boldsymbol{W}_1) -\mathrm{Tr}( \boldsymbol{W}_{2}^{H}\boldsymbol{S}_a\boldsymbol{W}_2 ) | \nonumber
\\
&=| \mathrm{Tr}( ( \boldsymbol{W}_1-\boldsymbol{W}_2 ) ^H\boldsymbol{S}_a\boldsymbol{W}_1 ) -\mathrm{Tr}( \boldsymbol{W}_{2}^{H}\boldsymbol{S}_a( \boldsymbol{W}_1-\boldsymbol{W}_2 ) ) | \nonumber
\\
&\le \lVert \boldsymbol{W}_1-\boldsymbol{W}_2 \rVert _F\lVert \boldsymbol{S}_a \rVert _2( \lVert \boldsymbol{W}_1 \rVert _F+\lVert \boldsymbol{W}_2 \rVert _F ), 
\label{eq:proof_trace_inequality_1}
\\
&| \mathrm{Tr}( \boldsymbol{W}_{1}^{H}\boldsymbol{S}_a\boldsymbol{W}_1 ) -\bar{\mathcal{P}}_{\text{d},a} | 
\le \lVert \boldsymbol{S}_a \rVert _2\lVert \boldsymbol{W}_1 \rVert _{F}^{2}+\bar{\mathcal{P}}_{\text{d},a},
\label{eq:proof_trace_inequality_2}
\end{align}
where the power limitation holds
\begin{align}
\label{eq:proof_power_inequality}
&\lVert \boldsymbol{W}_1 \rVert _F\le \sqrt{P_\text{B}}, \quad \lVert \boldsymbol{W}_2 \rVert _F\le \sqrt{P_\text{B}}.
\end{align}
Substituting \eqref{eq:proof_trace_inequality_1}, \eqref{eq:proof_trace_inequality_2}, and \eqref{eq:proof_power_inequality} into \eqref{eq:proof_norm_inequality}, we obtain
\begin{align}
&\lVert \nabla g_a\left( \boldsymbol{W}_1 \right) -\nabla g_a\left( \boldsymbol{W}_2 \right) \rVert  \nonumber
\\
&\le 4\lVert \boldsymbol{S}_a \rVert _2\left( P_\text{B}\lVert \boldsymbol{S}_a \rVert _2+\bar{\mathcal{P}}_{\text{d},a} \right) \lVert \boldsymbol{W}_1-\boldsymbol{W}_2 \rVert _F \nonumber
\\
&\quad +4\lVert \boldsymbol{S}_a \rVert _2( 2\sqrt{P_\text{B}}\lVert \boldsymbol{S}_a \rVert _2\lVert \boldsymbol{W}_1-\boldsymbol{W}_2 \rVert _F ) \sqrt{P_\text{B}} \nonumber
\\
&=\left( 12P_\text{B}\lVert \boldsymbol{S}_a \rVert _{2}^{2}+4\lVert \boldsymbol{S}_a \rVert _2\bar{\mathcal{P}}_{\text{d},a} \right) \lVert \boldsymbol{W}_1-\boldsymbol{W}_2 \rVert _F,
\end{align}
which is exactly the definition of the Lipschitz constant. \end{IEEEproof}
Consequently, at each iteration, replacing each $g_a(\boldsymbol{W})$ by its concave upper bound $\mathcal{U}_a( \boldsymbol{W}\mid \boldsymbol{W}^{( t )} )$ on $\boldsymbol {W} ^ {(t)}$ in \eqref{eq:W_sensing_bound} provides a tight lower bound to maximize the objective function $-\rho \text{NMSE}(\boldsymbol{W})$ of \eqref{p2:W_nmse_obj}, guaranteeing non-decreasing objective values in updates to $\boldsymbol{W}$ with a standard MM step.

\subsubsection{Communication SINR term with LDT and QT}
Introducing auxiliary real-valued variables $\{\mu_k>0,\forall k\}$, we remove the fractional term from the logarithmic function using the LDT \cite{ref27}
\begin{align}
&\log_2 \Big( 1+\frac{| \boldsymbol{f}_{k}^{H}\boldsymbol{w}_{k}^{} |^2}{\sum_{i=1,i\ne k}^{K+M}{| \boldsymbol{f}_{k}^{H}\boldsymbol{w}_{i}^{} |^2}+\sigma _{k}^{2}} \Big) \nonumber
\\
&=\underset{\mu _k\ge 0}{\max}\,\log_2( 1\!+\!\mu _k) \!-\!\frac{\mu _k}{\ln 2}\!+\!\frac{1+\mu _k}{\ln 2}\frac{| \boldsymbol{f}_{k}^{H}\boldsymbol{w}_{k}^{}|^2}{\sum_{i=1}^{K+M}{| \boldsymbol{f}_{k}^{H}\boldsymbol{w}_{i}^{} |^2}\!+\!\sigma _{k}^{2}}.
\end{align}
Next, introducing complex-valued variables $\{\eta_k,\forall k\}$,  we apply QT to transform the fractional SINR term into
\begin{align}
&\frac{|\boldsymbol{f}_{k}^{H}\boldsymbol{w}_{k}^{}|^2}{\sum_{i=1}^{K+M}{|\boldsymbol{f}_{k}^{H}\boldsymbol{w}_{i}^{}|^2}+\sigma _{k}^{2}} \nonumber
\\
&=\underset{\eta_k\in \mathbb{C}}{\max}\,\,2\text{Re}\{ \eta_{k}^{*}\boldsymbol{f}_{k}^{H}\boldsymbol{w}_{k}^{} \} \!-\!\left| \eta_k \right|^2\Big(\! \sum_{i=1}^{K+M}\!{|\boldsymbol{f}_{k}^{H}\boldsymbol{w}_{i}^{}|^2}\!+\!\sigma _{k}^{2} \Big). \!\!
\end{align}
These auxiliary variables can be updated in closed form
\begin{align}
&\mu _{k}^{*}=\frac{|\boldsymbol{f}_{k}^{H}\boldsymbol{w}_{k}^{}|^2}{\sum_{i=1,i\ne k}^{K+M}{|\boldsymbol{f}_{k}^{H}\boldsymbol{w}_{i}^{}|^2}+\sigma _{k}^{2}},
\label{eq:QT_update_1}
\\
&\eta_{k}^{*}=\frac{\boldsymbol{f}_{k}^{H}\boldsymbol{w}_{k}^{}}{\sum_{i=1}^{K+M}{|\boldsymbol{f}_{k}^{H}\boldsymbol{w}_{i}^{}|^2}+\sigma _{k}^{2}}.
\label{eq:QT_update_2}
\end{align}
Consequently, the communication sum rate is equivalently transformed into a non-fractional quadratic function of $\boldsymbol{W}$
\begin{align}
\!&\mathcal{Y}\left( \boldsymbol{W}\mid \mu _{k}^{},\eta_{k}^{} \right) \nonumber
\\
&\!=\!\frac{1\!\!+\!\mu _k}{\ln 2}\Big( 2\text{Re}\{\eta_{k}^{*}\boldsymbol{f}_{k}^{H}\boldsymbol{w}_{k}^{}\}\!-\!\left| \eta_k \right|^2\!\Big( \!\sum_{i=1}^{K+M}{\!\!}|\boldsymbol{f}_{k}^{H}\boldsymbol{w}_{i}^{}|^2\!\!+\!\sigma _{k}^{2} \Big) \!\Big). \!\!\!
\label{eq:QT_expression}
\end{align}

\subsubsection{Closed form solution for the QCQP}
With $\{\mu_{k},\eta_{k}\}$ fixed and $g_{a}$ replaced by $\mathcal{U}_{a}$, (P2) reduces to the convex quadratic constrained quadratic program (QCQP) stated as below
\begin{align}
\label{p2.1:objective}
\text{(P2.1):} \underset{\boldsymbol{W}}{\max}\,&\sum_{k=1}^K{\mathcal{Y}\left( \boldsymbol{W}\!\mid\! \mu_{k},\eta_{k} \right)}-\frac{\rho}{D} \sum_{a=1}^A{\mathcal{U}_a( \boldsymbol{W}\!\mid\! \boldsymbol{W}^{(t)})}  
\\
\text{s.t.}\,\,&\eqref{p1:power}. \nonumber
\end{align}
In general, (P2.1) can be solved using convex optimization tools. However, we derive a solution based on Lagrange duality and the Karush-Kuhn-Tucker (KKT) conditions. Specifically, at the current $\boldsymbol{W}^{(t)}$, the objective \eqref{p2.1:objective} can be compactly rewritten as a summation with a linear and a quadratic term
\begin{align}
\underset{\boldsymbol{W}}{\max}\,\, &-\mathrm{Tr}( \boldsymbol{W}^H\boldsymbol{QW})+2\text{Re}\{ \mathrm{Tr}( \boldsymbol{P}^H\boldsymbol{W} ) \} 
\\
\text{s.t.}& \,\, \text{\eqref{p1:power}}, \nonumber
\end{align}
where we omit constant values that do not affect optimization, and the matrices $\boldsymbol{Q}$ and $\boldsymbol{P}$ are defined respectively by
\begin{subequations}
\label{eq:Q_P}
\begin{align}
&\boldsymbol{Q}=\boldsymbol{Q}_{\text{C}}+\rho \boldsymbol{Q}_{\text{S}}, 
\\
&\boldsymbol{P}=\boldsymbol{P}_{\text{C}}+\rho \boldsymbol{P}_{\text{S}},
\end{align}
\end{subequations}
with equivalent expressions from \eqref{eq:W_sensing_bound} and \eqref{eq:QT_expression} as
\begin{subequations}
\label{eq:QC_QS_PC_PS}
\begin{align}
&\boldsymbol{Q}_{\text{C}}=\frac{1}{\ln 2}\sum_{k=1}^K{\left( 1+\mu _k \right) \left| \eta_k \right|^2\boldsymbol{f}_k\boldsymbol{f}_{k}^{H}},
\\
&\boldsymbol{Q}_{\text{S}}=\frac{1}{2D}\sum_{a=1}^A{L_{a}^{\text{lip}}}\boldsymbol{I}_{M},
\\
&\boldsymbol{P}_{\text{C}}=\frac{1}{\ln 2}\left[ \left( 1+\mu _1 \right) \eta_{1}^{}\boldsymbol{f}_{1}^{},\dots ,\left( 1+\mu _K \right) \eta_{K}^{}\boldsymbol{f}_{K}^{} \right], 
\\
&\boldsymbol{P}_{\text{S}}=\frac{1}{2D}\sum_{a=1}^A{L_{a}^{\text{lip}}\boldsymbol{W}^{\left( t \right)}-\boldsymbol{G}_{a}\big( \boldsymbol{W}^{(t)} \big)}.
\end{align}
\end{subequations}
The corresponding Lagrangian with a multiplier $\nu>0$ is
\begin{align}
\mathcal{L}( \boldsymbol{W},\nu ) =&-\mathrm{Tr}( \boldsymbol{W}^H\boldsymbol{QW} )+2\text{Re}\{ \mathrm{Tr}( \boldsymbol{P}^H\boldsymbol{W} )\} \nonumber
\\
&-\nu ( \lVert \boldsymbol{W} \rVert _{F}^{2}-P_{\text{B}} ), 
\end{align}
with KKT conditions for stationarity, complementary slackness, primal feasibility, and dual feasibility, respectively, being
\begin{subequations}
\begin{align}
&-(\boldsymbol{Q}+\nu ^*\boldsymbol{I})\boldsymbol{W}^*+\boldsymbol{P}=0,
\\
&\nu ^*\left( \lVert \boldsymbol{W}^* \rVert _{F}^{2}-P_{\text{B}} \right) =0,
\\
&\lVert \boldsymbol{W}^* \rVert _{F}^{2}\le P_{\text{B}},
\\
&\nu ^*\ge 0,
\end{align}
\end{subequations}
 which yields the water-filling-like solution
\begin{align}
\label{eq:W_optimal}
\boldsymbol{W}\left( \nu \right) =\left( \boldsymbol{Q}+\nu \boldsymbol{I} \right) ^{-1}\boldsymbol{P}, \nu \ge 0,
\end{align}
where $\nu$ is found by bisection search to satisfy the power constraint \eqref{p1:power} with equality. Since $\nu$ enters as a scalar shift on the identity, we precompute the eigenvalue decomposition (EVD) $\boldsymbol{Q}=\boldsymbol{U\varLambda U}^H$ with $\boldsymbol{\varLambda }=\mathrm{diag}( \lambda _1,\dots ,\lambda _M)$ and $\lambda_m$ being the eigenvalue; thus, for $\nu >-\min \{\lambda_m\}_{m=1}^{M}$, we have
\begin{subequations}
\label{eq:bisection_1}
\begin{align}
\boldsymbol{W}\left( \nu \right) &=\boldsymbol{U}\left( \boldsymbol{\varLambda }+\nu \boldsymbol{I} \right) ^{-1}\boldsymbol{U}^H\boldsymbol{P}
\\
&=\boldsymbol{U}\mathrm{diag}\left( \left[ \frac{1}{\lambda _1+\nu},\dots ,\frac{1}{\lambda _M+\nu} \right] \right) \boldsymbol{U}^H\boldsymbol{P},
\end{align}
\end{subequations}
so each bisection step costs only a diagonal scaling, without refactorization, and evaluates
\begin{align}
\label{eq:bisection_2}
&\lVert \boldsymbol{W}\left( \nu \right) \rVert _{F}^{2}=\sum_{i=1}^M{\frac{\lVert [ \boldsymbol{U}^H\boldsymbol{P} ] _{i,:} \rVert _{2}^{2}}{( \lambda _i+\nu ) ^2}},
\end{align}
where $[\cdot]_{i,:}$ denote the $i$-th row of a matrix.

\subsubsection{Overall algorithm and complexity analysis for subproblem 1} The overall low-complexity algorithm for solving (P2) is summarized in Algorithm 1. Each inner iteration performs one EVD of an $M{\times}M$ Hermitian matrix at cost $\mathcal{O}(M^3)$ and a scalar bisection whose per-step cost is $\mathcal{O}(M(K+M))$ to calculate $\|\boldsymbol W(\nu)\|_{F}^{2}$; the number of bisection steps grows only logarithmically with precision. The MM construction provides a tight lower bound of the objective in (P2), which is non-decreasing over $\boldsymbol W$-updates. Under the power ball, the sequence $\{\boldsymbol W^{(t)}\}$ admits limit points that satisfy the first-order stationarity conditions of the surrogate built from MM and QT.

\begin{algorithm}[t]
\caption{Low-complexity transmit beamforming based on QT, MM, and bisection search for (P2)}\label{alg:QTMM}
\small
\begin{algorithmic}[1]
\STATE \textbf{Input:} $\{\boldsymbol f_k\}_{k=1}^{K}$, $\{\boldsymbol f_{\text{S},a}\}_{a=1}^{A}$, $\{\mathcal{P}_{\text{d},a}\}$, $P_{\text{B}}$, $\rho$.
\STATE \textbf{Initialization:} $\boldsymbol W^{(0)}$, using zero forcing in the first outer iterate, or using previous outer iterate results; Set $t=0$.
\REPEAT
\STATE Compute $\|\boldsymbol{S}_a\|_2^2$ via \eqref{eq:S_a},  $L_a^{\text{lip}}$ via \eqref{eq:L_a}, $\boldsymbol G_a(\boldsymbol W^{(t)})$ via \eqref{eq:G_a}, and $\mathcal{U}(\boldsymbol W\mid\boldsymbol W^{(t)})$ via \eqref{eq:W_sensing_bound}.
\STATE Build $\boldsymbol{Q}_{\text{C}}$, $\boldsymbol{P}_{\text{C}}$, $\boldsymbol{Q}_{\text{S}}$, and $\boldsymbol{P}_{\text{S}}$ from \eqref{eq:QC_QS_PC_PS}. Set $\boldsymbol{Q}$ and $\boldsymbol{P}$ via \eqref{eq:Q_P}. 
\STATE Perform EVD by $\boldsymbol Q=\boldsymbol U\boldsymbol\Lambda\boldsymbol U^{H}$. Find $\nu^{\star}$ by bisection using \eqref{eq:bisection_1} and \eqref{eq:bisection_2}; set $\boldsymbol W^{(t{+}1)}=\boldsymbol W(\nu^{\star})$ via \eqref{eq:W_optimal}.
\STATE Update auxiliaries $\mu_k^{(t{+}1)}$ by \eqref{eq:QT_update_1} and $\eta_k^{(t{+}1)}$ by \eqref{eq:QT_update_2}.
\STATE $t\leftarrow t{+}1$.
\UNTIL{the relative increment on the objective value of \eqref{p2.1:objective} or $\|\boldsymbol W^{(t)}{-}\boldsymbol W^{(t-1)}\|_F/\|\boldsymbol W^{(t-1)}\|_F$ below a predetermined threshold $\varepsilon$, or $t$ hits the maximum number of iterations $t_{\max}$.}
\STATE \textbf{Output:} $\boldsymbol W^{\star}=\boldsymbol W^{(t)}$.
\end{algorithmic}
\end{algorithm}

\subsection{Subproblem 2: Passive Beamforming}

Given fixed $\boldsymbol{W}$ and $(\boldsymbol{r}^{\text{B}},\boldsymbol{r}^{\text{R}})$, we optimize $\boldsymbol{\theta}$ via 
\begin{align}
\label{p3:objective}
\text{(P3)} :\underset{\boldsymbol{\theta}}{\max}\,\,&\sum_{k=1}^K{\log _2\left( 1 +\text{SINR}_k(\boldsymbol{\theta}) \right)} - \rho\text{NMSE}(\boldsymbol\theta) 
\\
\text{s.t.} \quad & \eqref{p1:phase}. \nonumber
\end{align}
The unique challenge in RIS passive beamforming is the unit-modulus constraint. Nevertheless, it inherently defines a Riemannian manifold, known as the \textit{complex circle manifold}
\begin{align}
\mathcal{R}_{\boldsymbol{\theta}} = \left\{ \boldsymbol{\theta} \in \mathbb{C}^N : |\theta_n| = 1, \forall n \in \mathcal{N} \right\}, 
\end{align}
which enables the use of manifold optimization techniques to efficiently exploit the smoothness of objective functions and the structure of manifold feasible regions.

\subsubsection{Problem reformulation} 
First, we rewrite expressions in \eqref{p3:objective} with respect to $\boldsymbol{\theta}$, and recast an unconstrained manifold optimization problem. Specifically, for the $a$-th angular sample, $\mathcal{P}_a(\vartheta,\varphi;\boldsymbol{\theta})$ in \eqref{eq:transmit_beam_pattern} expands as a quadratic form
\begin{align}
\mathcal{P}_a(\vartheta,\varphi;\boldsymbol{\theta})
&=\boldsymbol{\theta}^H\boldsymbol{\Xi}_{\text{S},a}\boldsymbol{\theta}
+2\Re\{\boldsymbol{\xi}_{\text{S},a}^H\boldsymbol{\theta}\}+\xi_{\text{S},a},
\end{align}
where $\boldsymbol{\Xi}_{\text{S},a}$, $\boldsymbol{\xi}_{\text{S},a}$, and $\xi_{\text{S},a}$ are computed based on \eqref{eq:channel_BS_target} by
\begin{subequations}
\begin{align}
\label{eq:Xi}
&\!\boldsymbol{\Xi}_{\mathrm{S},a}=G^{\text{R}}(\boldsymbol{r}^{\text{R}})
\mathrm{diag}(\boldsymbol{t}_a^{\mathrm{RT}})
\boldsymbol{B}^H\boldsymbol{W}\boldsymbol{W}^H\boldsymbol{B}
\mathrm{diag}(\boldsymbol{t}_a^{\mathrm{RT}})^H,\!\!
\\
&\!\boldsymbol{\xi}_{\mathrm{S},a}^H=\!\sqrt{G^{\text{B}}(\boldsymbol{r}^{\text{B}})G^{\text{R}}(\boldsymbol{r}^{\text{R}})}
(\boldsymbol{t}_a^{\mathrm{BT}})^H\boldsymbol{W}\boldsymbol{W}^H\boldsymbol{B}\mathrm{diag}(\boldsymbol{t}_a^{\mathrm{RT}}),\!\!\!
\\
&\!\xi_{\mathrm{S},a}=G^{\text{B}}(\boldsymbol{r}^{\text{B}})(\boldsymbol{t}_a^{\mathrm{BT}})^H\boldsymbol{W}\boldsymbol{W}^H\boldsymbol{t}_a^{\mathrm{BT}}.\!\!
\end{align}
\end{subequations}
Similarly, the signal or interference term of any beam $\boldsymbol w_i$ for the $k$-th communication user in the SINR expression with effective downlink channel $\boldsymbol f_k(\boldsymbol\theta)$ can be written as
\begin{align}
\big|\boldsymbol{f}_k^H(\boldsymbol\theta)\boldsymbol{w}_i\big|^2
&=\boldsymbol{\theta}^H\boldsymbol{\Xi}_{\mathrm{C},k,i}\boldsymbol{\theta}
+2\Re\{\boldsymbol{\xi}_{\mathrm{C},k,i}^H\boldsymbol{\theta}\}
+\xi_{\mathrm{C},k,i},
\end{align}
with $\boldsymbol{\Xi}_{\text{C},k,i}$, $\boldsymbol{\xi}_{\text{C},k,i}$, and $\xi_{\text{C},k,i}$ computed based on \eqref{eq:channel_BS_user}
\begin{subequations}
\begin{align}
\boldsymbol{\Xi}_{\mathrm{C},k,i}&=\mathrm{diag}(\boldsymbol{g}_k^*)\,\boldsymbol{B}^H\boldsymbol{w}_i\boldsymbol{w}_i^H\boldsymbol{B}\,\mathrm{diag}(\boldsymbol{g}_k),\\
\boldsymbol{\xi}_{\mathrm{C},k,i}^H&=\boldsymbol{h}_k^H\boldsymbol{w}_i\boldsymbol{w}_i^H\boldsymbol{B}\,\mathrm{diag}(\boldsymbol{g}_k),\\
\xi_{\mathrm{C},k,i}&=\boldsymbol{h}_k^H\boldsymbol{w}_i\boldsymbol{w}_i^H\boldsymbol{h}_k.
\end{align}
\end{subequations}
Define real-valued scalars $S_a(\boldsymbol{\theta})$, $C_{1,k}(\boldsymbol{\theta})$, and $C_{2,k}(\boldsymbol{\theta})$ as
\begin{subequations}
\begin{align}
\!\!\!S_a(\boldsymbol{\theta})&=\boldsymbol{\theta}^H\boldsymbol{\Xi}_{\mathrm{S},a}\boldsymbol{\theta}
+2\Re\{\boldsymbol{\xi}_{\mathrm{S},a}^H\boldsymbol{\theta}\}
+\xi_{\mathrm{S},a}-\bar{\mathcal{P}}_{\mathrm{d},a},
\\
\!\!\!C_{1,k}(\boldsymbol{\theta})&=\boldsymbol{\theta}^H\boldsymbol{\Xi}_{\mathrm{C},k,k}\boldsymbol{\theta}
+2\Re\{\boldsymbol{\xi}_{\mathrm{C},k,k}^H\boldsymbol{\theta}\}
+\xi_{\mathrm{C},k,k},
\\
\!\!\!C_{2,k}(\boldsymbol{\theta})&\!\!=\!\!\sum_{i=1,i\neq k}^{K+M}\!\!\!\Big(\!
\boldsymbol{\theta}^H\boldsymbol{\Xi}_{\mathrm{C},k,i}\boldsymbol{\theta}
\!+\!2\Re\{\boldsymbol{\xi}_{\mathrm{C},k,i}^H\boldsymbol{\theta}\!\}
\!+\!\xi_{\mathrm{C},k,i}\!\Big)\!\!+\!\sigma_k^2, \!\!\!
\end{align}
\end{subequations}
we can rewrite the objective function in \eqref{p3:objective} as
\begin{align}
F(\boldsymbol\theta)
= \sum_{k=1}^{K}\log_2\!\left(1+\frac{C_{1,k}(\boldsymbol\theta)}{C_{2,k}(\boldsymbol\theta)}\right)
-\frac{\rho}{D}\sum_{a=1}^{A}\bigl|S_a(\boldsymbol\theta)\bigr|^2,
\end{align}
which further arrives at a standard Riemannian unconstrained manifold optimization problem stated as
\begin{align}
\text{(P3.1):}\,\,\min_{\boldsymbol{\theta}\in\mathcal{R}_{\boldsymbol{\theta}}}&\,\,-F\left( \boldsymbol{\theta } \right). \nonumber
\end{align}

\subsubsection{RCG method}
To solve (P3.1), we adopt the RCG method, which generalizes the classical conjugate gradient algorithm over curved spaces, as a manifold is a topological space that locally resembles the Euclidean space \cite{ref28,ref29}. At any point, tangent vectors give all feasible updating directions, and their set is the tangent space
\begin{align}
T_{\boldsymbol{\theta}}=\big\{\boldsymbol{\varsigma}\in\mathbb{C}^N:\;\Re\{\boldsymbol{\varsigma}\odot \boldsymbol{\theta}^*\}=\boldsymbol{0}\big\}.
\end{align}
Based on this definition, the RCG method proceeds by i) computing the Riemannian gradient, ii) determining the conjugate search direction, and iii) applying a retraction operation to ensure that updated variables remain on the manifold.

The Riemannian gradient is a vector field on the manifold obtained by projecting the Euclidean gradient onto the tangent space. Using Wirtinger derivatives calculus for the real-valued $F(\boldsymbol{\theta})$ with all $\boldsymbol{\xi}_{\text{S},a}$ and $\boldsymbol{\xi}_{\text{C},k,i}$ being Hermitian, we obtain
\begin{align}
\nabla_{\boldsymbol{\theta}} S_a(\boldsymbol{\theta})
&=2\boldsymbol{\Xi}_{\mathrm{S},a}\boldsymbol{\theta}+2\boldsymbol{\xi}_{\mathrm{S},a},\\
\nabla_{\boldsymbol{\theta}} C_{1,k}(\boldsymbol{\theta})
&=2\boldsymbol{\Xi}_{\mathrm{C},k,k}\boldsymbol{\theta}+2\boldsymbol{\xi}_{\mathrm{C},k,k},\\
\nabla_{\boldsymbol{\theta}} C_{2,k}(\boldsymbol{\theta})
&=\sum_{i=1,\,i\neq k}^{K+M}\big(2\boldsymbol{\Xi}_{\mathrm{C},k,i}\boldsymbol{\theta}+2\boldsymbol{\xi}_{\mathrm{C},k,i}\big).
\end{align}
Using the chain rule, the complete Euclidean gradient expression $\nabla_{\boldsymbol{\theta}}F(\boldsymbol{\theta})$ can be computed by \eqref{eq:RCG_Egrad} at the top of the next page.
\begin{figure*}
\begin{subequations}
\label{eq:RCG_Egrad}
\begin{align}
&\nabla _{\boldsymbol{\theta }}F\left( \boldsymbol{\theta } \right) =\frac{1}{\ln 2}\sum_{k=1}^K{\frac{C_{2,k}\left( \boldsymbol{\theta } \right) \,\nabla _{\boldsymbol{\theta }}C_{1,k}\left( \boldsymbol{\theta } \right) -C_{1,k}\left( \boldsymbol{\theta } \right) \,\nabla _{\boldsymbol{\theta }}C_{2,k}\left( \boldsymbol{\theta } \right)}{C_{2,k}\left( \boldsymbol{\theta } \right) \left( C_{1,k}\left( \boldsymbol{\theta } \right) +C_{2,k}\left( \boldsymbol{\theta } \right) \right)}}- \frac{\rho}{D}\sum_{a=1}^A{2}S_a\left( \boldsymbol{\theta } \right) \,\nabla _{\boldsymbol{\theta }}S_a\left( \boldsymbol{\theta } \right) 
\\
&\qquad \,\,\,\quad =\frac{2}{\ln 2}\sum_{k=1}^K{\frac{C_{2,k}\left( \mathbf{\Xi }_{\text{C},k,k}\boldsymbol{\theta }+\boldsymbol{\xi }_{\text{C},k,k} \right) -C_{1,k}\sum_{i=1,i\ne k}^K{\mathbf{\Xi }_{\text{C},k,i}\boldsymbol{\theta }+\boldsymbol{\xi }_{\text{C},k,i}}}{C_{2,k}\left( C_{1,k}+C_{2,k} \right)}}-\frac{4\rho}{D}\sum_{a=1}^A{S_a}\left( \boldsymbol{\theta } \right) \left( \mathbf{\Xi }_{\text{S},a}\boldsymbol{\theta }+\boldsymbol{\xi }_{\text{S},a} \right) 
\\
& \overline{\ \ \ \ \ \ \ \ \ \ \ \ \ \ \ \ \ \ \ \ \ \ \ \ \ \ \ \ \ \ \ \ \ \ \ \ \ \ \ \ \ \ \ \ \ \ \ \ \ \ \ \ \ \ \ \ \ \ \ \ \ \ \ \ \ \ \ \ \ \ \ \ \ \ \ \ \ \ \ \ \ \ \ \ \ \ \ \ \ \ \ \ \ \ \ \ \ \ \ \ \ \ \ \ \ \ \ \ \ \ \ \ \ \ \ \ \ \ \ \ \ \ \ \ \ \ \ \ \ \ \ \ \ \ \ \ \ \ \ \ \ \ \ \ \ \ \ } \nonumber
\end{align}
\end{subequations}
\vspace{-36pt}
\end{figure*}
Then, the Riemannian gradient is obtained via the orthogonal projection $\mathsf{Proj}$ for the Euclidean vector $\nabla_{\boldsymbol{\theta}}F(\boldsymbol{\theta})$ onto the tangent space $T_{\boldsymbol{\theta}}$ as follows
\begin{subequations}
\label{eq:RCG_Rgrad}
\begin{align}
\mathsf{Rgrad}F\left( \boldsymbol{\theta } \right) &=\mathsf{Proj}_{\boldsymbol{\theta }}\left( \nabla _{\boldsymbol{\theta }}F\left( \boldsymbol{\theta } \right) \right)
\\
&=\nabla _{\boldsymbol{\theta }}F\left( \boldsymbol{\theta } \right) -\text{Re}\{\nabla _{\boldsymbol{\theta }}F\left( \boldsymbol{\theta } \right) \odot \boldsymbol{\theta }^*\}\odot \boldsymbol{\theta },
\end{align}
\end{subequations}
where \(\odot\) denotes the Hadamard product.
Next, in iteration $\jmath$, the descent direction $\boldsymbol{\tau }^{(\jmath)}$ for $\boldsymbol{\theta}^{(\jmath)}$ is determined using the Polak-Ribiere formula. For brevity, the function $F(\boldsymbol{\theta}^{(\jmath)})$ is abbreviated as $F^{(\jmath)}$ in the following. Let $\nabla _{\mathcal{R}_{\boldsymbol{\theta}}}F^{(\jmath)}$ denote the Riemannian gradient in iteration $\jmath$. The conjugate gradient update coefficient $\chi^{(\jmath)}$, which adjusts the new search direction accounting for the curvature of the gradient, is 
\begin{align}
\label{eq:RCG_conjugate}
&\chi^{\left( \jmath \right)}=\frac{\left< \nabla _{\mathcal{R}_{\boldsymbol{\theta }}}F^{\left( \jmath \right)},\nabla _{\mathcal{R}_{\boldsymbol{\theta }}}F^{\left( \jmath \right)}-\nabla _{\mathcal{R}_{\boldsymbol{\theta }}}F^{\left( \jmath-1 \right)} \right>}{\left< \nabla _{\mathcal{R}_{\boldsymbol{\theta }}}F^{\left( \jmath-1 \right)},\nabla _{\mathcal{R}_{\boldsymbol{\theta }}}F^{\left( \jmath-1 \right)} \right>},
\end{align}
where $\left\langle \cdot, \cdot \right\rangle$ denotes the Riemannian metric, being Euclidean inner product here. The descent direction is then updated as
\begin{align}
\label{eq:RCG_direction}
&\boldsymbol{\tau}^{(\jmath)} = -\nabla _{\mathcal{R}_{\boldsymbol{\theta }}}F^{\left( \jmath \right)}+ \chi^{(\jmath)} \mathsf{Proj}_{\boldsymbol{\theta}^{(\jmath)}}(\boldsymbol{\tau}^{(\jmath-1)}),
\end{align}
here the projection represents a simple vector transport operation. After updating the variable by moving along the descent direction, we map the new point back onto the manifold using the retraction operation, ensuing that the updated variables $\boldsymbol{\theta}^{(\jmath+1)}$ satisfy their manifold constraints, as follows
\begin{subequations}
\label{eq:RCG_retraction}
\begin{align}
\boldsymbol{\theta }^{\left( \jmath+1 \right)}&=\mathsf{Retr}_{\boldsymbol{\theta }^{\left( \jmath \right)}}( \varpi^{\left( \jmath \right)}\boldsymbol{\tau }^{\left(\jmath \right)} ) 
\\
&=\left[ \frac{( \boldsymbol{\theta }^{\left( \jmath \right)}+\varpi^{\left( \jmath \right)}\boldsymbol{\tau }^{\left( \jmath \right)}) _n}{| ( \boldsymbol{\theta }^{\left( \jmath \right)}+\varpi^{\left( \jmath \right)}\boldsymbol{\tau }^{\left( \jmath \right)} ) _n |} \right] ,
\end{align}
\end{subequations}
where $\varpi^{(\jmath)}$ is the step size determined by an Armijo backtracking line search method, and the element-wise normalization ensures that $\boldsymbol{\theta}^{(\jmath+1)}$ lies on the complex circle manifold.

\begin{algorithm}[t]
\caption{RCG for Passive RIS Beamforming (P3)}
\begin{algorithmic}[1]
\STATE \textbf{Input:} $\{\boldsymbol{\Xi}_{\text{S},a}, \boldsymbol{\xi}_{\text{S},a}, \xi_{\text{S},a}\}$ and $\{\boldsymbol{\Xi}_{\text{C},k,i}, \boldsymbol{\xi}_{\text{C},k,i}, \xi_{\text{C},k,i}\}$.
\STATE \textbf{Initialization:} $\boldsymbol{\theta}^{(0)}$ , $\boldsymbol{\tau}^{(0)}=-\mathsf{Rgrad}\,F(\boldsymbol{\theta}^{(0)})$.
\REPEAT
  \STATE Compute $\nabla_{\boldsymbol{\theta}}F(\boldsymbol{\theta}^{(\jmath)})$ via \eqref{eq:RCG_Egrad}) and $\mathsf{Rgrad}F(\boldsymbol{\theta})$ via \eqref{eq:RCG_Rgrad}.
  \STATE Update $\chi^{(\jmath)}$ via \eqref{eq:RCG_conjugate} and search direction $\boldsymbol{\tau}^{(\jmath)}$ via \eqref{eq:RCG_direction}.
  \STATE Armijo backtracking line search to get $\varpi^{(\jmath)}$.
  \STATE Perform retraction via \eqref{eq:RCG_retraction} and obtain  $\boldsymbol{\theta}^{(\jmath+1)}$.
  \STATE $\jmath \gets \jmath +1$
\UNTIL{Convergence: $\|\nabla_{\boldsymbol{\theta}}F(\boldsymbol{\theta}^{(\jmath)})\|/\sqrt{N}\le\varepsilon$ or $\jmath\ge \jmath_{\max}$.}
\STATE \textbf{Output:} $\boldsymbol{\theta}^{\star}=\boldsymbol{\theta}^{(\jmath)}$.
\end{algorithmic}
\end{algorithm} 

\subsubsection{Overall algorithm and complexity analysis for subproblem 2}
Algorithm 2 summarizes the manifold optimization procedure, guaranteeing convergence to a stationary point under standard regularity conditions [39]. The precomputation to build all $N\times N$ matrices $\{\boldsymbol{\varXi}_{\text{S},a}\}_{a=1}^{A}$ and $\{\boldsymbol{\varXi}_{\text{C},k,i}\}_{k\in[K],i\in[K+M]}$ scales as $\mathcal{O}(MN^2+MN(K+M)+(A+K(K+M))N^2)$. For each RCG iteration, including objective and Riemannian gradient evaluations, line search with a constant number of trials, costs $\mathcal{O}((A+K(K+M))N^2)$, since for every grid $a$ and pair $(k,i)$ one performs products of the form $\boldsymbol{\varXi}_{\text{C},k,i}\boldsymbol{\theta}$ or $\boldsymbol{\varXi}_{\text{S},a}\boldsymbol{\theta}$ and a few additional $\mathcal{O}(N^2)$ accumulations. The total computational complexity over $T_{\mathrm{RCG}}$ iterations is thus $\mathcal{O}(T_{\text{RCG}}(A+K(K+M))N^2)$ plus the aforementioned precomputation. These costs can be further reduced by avoiding explicit $N\times N$ formations and reusing the rank–1 decomposition for $\boldsymbol B^{H}\boldsymbol W\boldsymbol W^{H}\boldsymbol B$ in \eqref{eq:Xi}.

\subsection{Subproblem 3: Array Rotations}
Given $\boldsymbol{W}$ and $\boldsymbol{\theta}$ from the previous subproblems, we jointly optimize the array rotations $(\boldsymbol r^{\text{B}},\boldsymbol r^{\text{R}})$ of the BS and RIS via
\begin{align}
\label{p4:objective}
\text{(P4)} :\underset{\boldsymbol{r}^{\text{B}},\boldsymbol{r}^{\text{R}}}{\max}&\sum_{k=1}^K{\log _2\left( 1+\text{SINR}_k\left( \boldsymbol{r}^{\text{B}},\boldsymbol{r}^{\text{R}} \right) \right)} -\rho\text{NMSE}(\boldsymbol r^{\text{B}},\boldsymbol r^{\text{R}}) \!\!\!
\\
\text{s.t.} \quad & \eqref{p1:Bbox},\eqref{p1:Rbox} \nonumber
\end{align}
Due to the highly non-concave relationships involving the cosine and exponential functions in the directivity gain and steering vectors, respectively, in the objective function, and the relatively simple box constraints on the array rotation angles, we employ the projection gradient method (PGD) \cite{ref30}, an effective strategy for this nonlinear maximization problem with boundary value constraints.

\subsubsection{Gradient computation} 
The key step of PGD is to derive the gradient expression for the complex objective function \eqref{p4:objective}. We adopt $\boldsymbol{r}^{\text{A}}$ for any array rotation variable with $\text{A}\in\{\text{B,R}\}$ for brevity as defined in \eqref{eq:rotation_matrix_all}. Further, define a stacked six-dimensional vector $\boldsymbol{r}$ for all rotations as below
\begin{align}
\label{eq:stacked_rotation}
\boldsymbol{r}\triangleq [ (\boldsymbol{r}^{\text{B}})^T,( \boldsymbol{r}^{\text{R}})^T] ^T=\left[ r_{\mathrm{x}}^{\text{B}},r_{\mathrm{y}}^{\text{B}},r_{\mathrm{z}}^{\text{B}},r_{\mathrm{x}}^{\text{R}},r_{\mathrm{y}}^{\text{R}},r_{\mathrm{z}}^{\text{R}} \right] ^T\in \mathbb{R}^6.
\end{align}
The gradient of \eqref{p4:objective} w.r.t. the stacked rotation vector $\boldsymbol{r}$ is
\begin{align}
\frac{\partial F\left( \boldsymbol{r} \right)}{\partial \boldsymbol{r}}=&-\frac{2\rho}{D}\sum_{a=1}^A{\left( \mathcal{P}_a\left( \boldsymbol{r} \right) -\bar{\mathcal{P}}_{\text{d},a} \right) \frac{\partial \mathcal{P}_a\left( \boldsymbol{r} \right)}{\partial \boldsymbol{r}}} \nonumber
\\
&+\frac{1}{\ln 2}\sum_{k=1}^K{\frac{1}{1+\text{SINR}_k\left( \boldsymbol{r} \right)}\frac{\partial \text{SINR}_k\left( \boldsymbol{r} \right)}{\partial \boldsymbol{r}}},
\end{align}
where $\frac{\partial \text{SINR}_k(\boldsymbol{r})}{\partial \boldsymbol{r}}$ and $
\frac{\partial \mathcal{P}_a(\boldsymbol{r})}{\partial \boldsymbol{r}}$ are the two main computational branches. 
Define the signal and interference terms in the SINR expression by $C_{1,k}\left( \boldsymbol{r} \right) =|\boldsymbol{f}_{k}^{H}\left( \boldsymbol{r} \right) \boldsymbol{w}_k|^2$ and $C_{2,k}\left( \boldsymbol{r} \right) =\sum_{i\ne k}{|}\boldsymbol{f}_{k}^{H}\left( \boldsymbol{r} \right) \boldsymbol{w}_i|^2+\sigma _{k}^{2}$, respectively. Then we have
\begin{align}
\label{eq:dSINR}
\frac{\partial \,\text{SINR}_k\left( \boldsymbol{r} \right)}{\partial \boldsymbol{r}}=\frac{\tfrac{\partial C_{1,k}\left( \boldsymbol{r} \right)}{\partial \boldsymbol{r}}C_{2,k}\left( \boldsymbol{r} \right) -C_{1,k}\left( \boldsymbol{r} \right) \tfrac{\partial C_{2,k}\left( \boldsymbol{r} \right)}{\partial \boldsymbol{r}}}{\left( C_{2,k}\left( \boldsymbol{r} \right) \right) ^2},
\end{align}
with
\begin{subequations}
\label{eq:gradient_PGD_C}
\begin{align}
&\frac{\partial C_{1,k}( \boldsymbol{r} )}{\partial \boldsymbol{r}}=2\text{Re}\Big\{ \Big( \frac{\partial \boldsymbol{f}_k( \boldsymbol{r} )}{\partial \boldsymbol{r}} \Big) ^{H}\boldsymbol{w}_k\boldsymbol{w}_{k}^{H}\boldsymbol{f}_{k}( \boldsymbol{r}) \Big\} ,
\\
&\frac{\partial C_{2,k}( \boldsymbol{r})}{\partial \boldsymbol{r}}=\sum_{i\ne k}{2}\text{Re}\Big\{ \Big( \frac{\partial \boldsymbol{f}_k( \boldsymbol{r})}{\partial \boldsymbol{r}} \Big) ^{H}\boldsymbol{w}_i\boldsymbol{w}_{i}^{H}\boldsymbol{f}_{k}( \boldsymbol{r} ) \Big\}.
\end{align}
\end{subequations}
According to the effective user channel \eqref{eq:channel_BS_user}, we derive
\begin{subequations}
\begin{align}
&\frac{\partial \boldsymbol{f}_k( \boldsymbol{r})}{\partial \boldsymbol{r}^{\text{B}}}=\frac{\partial \boldsymbol{h}_k( \boldsymbol{r}^{\text{B}} )}{\partial \boldsymbol{r}^{\text{B}}}+\frac{\partial \boldsymbol{B}( \boldsymbol{r} )}{\partial \boldsymbol{r}^{\text{B}}}\mathrm{diag}( \boldsymbol{\theta } ) \boldsymbol{g}_k( \boldsymbol{r}^{\text{R}}) ,
\\
&\frac{\partial \boldsymbol{f}_k( \boldsymbol{r} )}{\partial \boldsymbol{r}^{\text{R}}}=\frac{\partial \boldsymbol{B}( \boldsymbol{r} )}{\partial \boldsymbol{r}^{\text{R}}}\,\mathrm{diag}( \boldsymbol{\theta } ) \boldsymbol{g}_k( \boldsymbol{r}^{\text{R}} ) +\boldsymbol{B}( \boldsymbol{r} ) \mathrm{diag}( \boldsymbol{\theta }) \frac{\partial \boldsymbol{g}_k( \boldsymbol{r}^{\text{R}} )}{\partial \boldsymbol{r}^{\text{R}}},
\end{align}
\end{subequations}
where $\frac{\partial \boldsymbol{h}_k( \boldsymbol{r}^{\text{B}} )}{\partial \boldsymbol{r}^{\text{B}}}$, $\frac{\partial \boldsymbol{g}_k( \boldsymbol{r}^{\text{R}} )}{\partial \boldsymbol{r}^{\text{R}}}$, $\frac{\partial \boldsymbol{B}( \boldsymbol{r} )}{\partial \boldsymbol{r}^{\text{R}}}$, and $\frac{\partial \boldsymbol{B}( \boldsymbol{r} )}{\partial \boldsymbol{r}^{\text{B}}}$ can be computed in \eqref{eq:gradient_PGD_h_g_B} at the top of the next page. 

Then we compute the derivatives of the steering vectors. We denote $\mathrm{X}\in\{\mathrm{BT},\mathrm{RT},\mathrm{BR},\mathrm{RB},\mathrm{BU},\mathrm{RU}\},
$ and omit the index for all the steering vectors with the same element-wise structure. For a generic steering vector,
\begin{align}
\boldsymbol{t}^{\text{X}}( \boldsymbol{r}^{\text{A}}) \triangleq \big[ e^{j\frac{2\pi}{\lambda}( \boldsymbol{u}^{\text{X}} ) ^T\boldsymbol{d}_{1}^{\text{A}}\left( \boldsymbol{r}^{\text{A}} \right)},...,e^{j\frac{2\pi}{\lambda}\left( \boldsymbol{u}^{\text{X}} \right) ^T\boldsymbol{d}_{J}^{\text{A}}( \boldsymbol{r}^{\text{A}} )} \big] ^T,
\end{align}
with element positions $\boldsymbol{d}_{v}^{\text{A}}\left( \boldsymbol{r}^{\text{A}} \right) =\boldsymbol{d}_{0}^{\text{A}}+\boldsymbol{R}\left( \boldsymbol{r}^{\text{A}} \right) \overline{\boldsymbol{d}}_{v}^{\text{A}}
$ as depicted in \eqref{eq:array_coordinate} and $v\in \{m,n\}$ denoting the index of BS or RIS element, the following relationships hold based on chain rules
\begin{subequations}
\label{eq:gradient_PGD_t}
\begin{align}
&\frac{\partial \boldsymbol{t}^{\text{X}}( \boldsymbol{r}^{\text{A}} )}{\partial \boldsymbol{r}^{\text{A}}}=\mathrm{diag}\left( \boldsymbol{\omega }^{\text{X}}\left( \boldsymbol{r}^{\text{A}} \right) \right) \,\boldsymbol{t}^{\text{X}}\left( \boldsymbol{r}^{\text{A}} \right) ,
\\
&\omega _{v}^{\text{X}}( \boldsymbol{r}^{\text{A}} ) =j\frac{2\pi}{\lambda}\left( \boldsymbol{u}^{\text{X}} \right) ^T\frac{\partial \boldsymbol{d}_{v}^{\text{A}}\left( \boldsymbol{r}^{\text{A}} \right)}{\partial \boldsymbol{r}^{\text{A}}},
\\
&\frac{\partial \boldsymbol{d}_{v}^{\text{A}}( \boldsymbol{r}^{\text{A}} )}{\partial \boldsymbol{r}^{\text{A}}}\!=\!\Big[ \frac{\partial \boldsymbol{R}( \boldsymbol{r}^{\text{A}} )}{\partial r_{\mathrm{x}}^{\text{A}}}\overline{\boldsymbol{d}}_{v}^{\text{A}},\frac{\partial \boldsymbol{R}( \boldsymbol{r}^{\text{A}} )}{\partial r_{\mathrm{y}}^{\text{A}}}\overline{\boldsymbol{d}}_{v}^{\text{A}},\frac{\partial \boldsymbol{R}( \boldsymbol{r}^{\text{A}})}{\partial r_{\mathrm{z}}^{\text{A}}}\overline{\boldsymbol{d}}_{v}^{\text{A}} \Big].\!\!\!
\end{align}
\end{subequations}

The derivatives of the rotation matrix $\boldsymbol{R}\left( \boldsymbol{r}^{\text{A}} \right) =\boldsymbol{R}_{\mathrm{x}}\left( r_{\mathrm{x}}^{\text{A}} \right) \boldsymbol{R}_{\mathrm{y}}\left( r_{\mathrm{y}}^{\text{A}} \right) \boldsymbol{R}_{\mathrm{z}}\left( r_{\mathrm{z}}^{\text{A}} \right) 
$ w.r.t each rotation variable are
\begin{subequations}
\begin{align}
&\frac{\partial \boldsymbol{R}( \boldsymbol{r}^{\text{A}} )}{\partial r_{\mathrm{x}}^{\text{A}}}=\frac{\partial \boldsymbol{R}_{\mathrm{x}}( r_{\mathrm{x}}^{\text{A}} )}{\partial r_{\mathrm{x}}^{\text{A}}}\boldsymbol{R}_{\mathrm{y}}( r_{\mathrm{y}}^{\text{A}} ) \boldsymbol{R}_{\mathrm{z}}( r_{\mathrm{z}}^{\text{A}} ) ,
\\
&\frac{\partial \boldsymbol{R}( \boldsymbol{r}^{\text{A}} )}{\partial r_{\mathrm{y}}^{\text{A}}}=\boldsymbol{R}_{\mathrm{x}}( r_{\mathrm{x}}^{\text{A}} ) \frac{\partial \boldsymbol{R}_{\mathrm{y}}( r_{\mathrm{y}}^{\text{A}} )}{\partial r_{\mathrm{y}}^{\text{A}}}\boldsymbol{R}_{\mathrm{z}}( r_{\mathrm{z}}^{\text{A}} ) ,
\\
&\frac{\partial \boldsymbol{R}( \boldsymbol{r}^{\text{A}} )}{\partial r_{\mathrm{z}}^{\text{A}}}=\boldsymbol{R}_{\mathrm{x}}( r_{\mathrm{x}}^{\text{A}} ) \boldsymbol{R}_{\mathrm{y}}( r_{\mathrm{y}}^{\text{A}} ) \frac{\partial \boldsymbol{R}_{\mathrm{z}}( r_{\mathrm{z}}^{\text{A}} )}{\partial r_{\mathrm{z}}^{\text{A}}},
\end{align}
\end{subequations}
and according to the definition in \eqref{eq:rotation_matrix}, we have
\begin{subequations}
\begin{align}
&\frac{\partial \boldsymbol{R}_{\mathrm{x}}\left( r_{\mathrm{x}}^{\text{A}} \right)}{\partial r_{\mathrm{x}}^{\text{A}}}=\left[ \begin{matrix}
	0&		0&		0\\
	0&		-\sin r_{\mathrm{x}}^{\text{A}}&		-\cos r_{\mathrm{x}}^{\text{A}}\\
	0&		\cos r_{\mathrm{x}}^{\text{A}}&		-\sin r_{\mathrm{x}}^{\text{A}}\\
\end{matrix} \right] ,
\\
&\frac{\partial \boldsymbol{R}_{\mathrm{y}}\left( r_{\mathrm{y}}^{\text{A}} \right)}{\partial r_{\mathrm{y}}^{\text{A}}}=\left[ \begin{matrix}
	-\sin r_{\mathrm{y}}^{\text{A}}&		0&		\cos r_{\mathrm{y}}^{\text{A}}\\
	0&		0&		0\\
	-\cos r_{\mathrm{y}}^{\text{A}}&		0&		-\sin r_{\mathrm{y}}^{\text{A}}\\
\end{matrix} \right] ,
\\
&\frac{\partial \boldsymbol{R}_{\mathrm{z}}\left( r_{\mathrm{z}}^{\text{A}} \right)}{\partial r_{\mathrm{z}}^{\text{A}}}=\left[ \begin{matrix}
	-\sin r_{\mathrm{z}}^{\text{A}}&		-\cos r_{\mathrm{z}}^{\text{A}}&		0\\
	\cos r_{\mathrm{z}}^{\text{A}}&		-\sin r_{\mathrm{z}}^{\text{A}}&		0\\
	0&		0&		0\\
\end{matrix} \right] .
\end{align}
\end{subequations}

For the directional gains and their derivatives, let \(\bar{\boldsymbol{n}}=[0,0,1]^T\) and \(\boldsymbol{n}(\boldsymbol{r}^{\text{A}})=\boldsymbol{R}(\boldsymbol{r}^{\text{A}})\bar{\boldsymbol{n}}\) be the array boresight.
According to the practical directivity model \eqref{eq:directivity_gain}, we have
\begin{subequations}
\begin{align}
&\frac{\partial \sqrt{G^{\text{A}}}}{\partial \boldsymbol{r}^{\text{A}}}=\frac{1}{2\sqrt{G^{\text{A}}}}\frac{\partial G^{\text{A}}}{\partial \boldsymbol{r}^{\text{A}}},
\\
&\frac{\partial G^{\text{A}}}{\partial \boldsymbol{r}^{\text{A}}}=\begin{cases}
	G_{0}^{\text{A}}\,b^{\text{A}}\bigl( \boldsymbol{n}^{T}(\boldsymbol{r}^{\text{A}})\boldsymbol{u} \bigr) ^{b^{\text{A}}-1}\bigl( \frac{\partial \boldsymbol{n}^{T}(\boldsymbol{r}^{\text{A}})}{\partial \boldsymbol{r}^{\text{A}}}\boldsymbol{u} \bigr) ,\\
	\qquad \qquad \quad \quad \,\,       \text{when } \boldsymbol{n}^{T}(\boldsymbol{r}^{\text{A}})\boldsymbol{u}>0,\\
	0, \,\, \quad \qquad \qquad        \text{when } \boldsymbol{n}^{T}(\boldsymbol{r}^{\text{A}})\boldsymbol{u}\le 0.\\
\end{cases}\quad 
\\
&\frac{\partial \boldsymbol{n}^T}{\partial \boldsymbol{r}^{\text{A}}}=\bar{\boldsymbol{n}}^T\frac{\partial \boldsymbol{R}(\boldsymbol{r}^{\text{A}})}{\partial \boldsymbol{r}^{\text{A}}}.
\end{align}
\end{subequations}
At the non-differentiable boundary, subgradients are used.

\begin{figure*}
\begin{subequations}
\label{eq:gradient_PGD_h_g_B}
\begin{align}
&\,\,\frac{\partial \boldsymbol{h}_k( \boldsymbol{r}^{\text{B}} )}{\partial \boldsymbol{r}^{\text{B}}}=\sum_{l=1}^{L_k}{\delta_{k,l}\bigg[ \frac{\partial \sqrt{G_{k,l}^{\text{B}}( \boldsymbol{r}^{\text{B}} )}}{\partial \boldsymbol{r}^{\text{B}}}\boldsymbol{t}_{k,l}^{\text{BU}}( \boldsymbol{r}^{\text{B}} ) +\sqrt{G_{k,l}^{\text{B}}( \boldsymbol{r}^{\text{B}} )}\frac{\partial \boldsymbol{t}_{k,l}^{\text{BU}}( \boldsymbol{r}^{\text{B}} )}{\partial \boldsymbol{r}^{\text{B}}} \bigg]},
\\
&\,\,\frac{\partial \boldsymbol{g}_k( \boldsymbol{r}^{\text{R}} )}{\partial \boldsymbol{r}^{\text{R}}}=\sum_{q=1}^{Q_k}{\psi_{k,q}\bigg[ \frac{\partial \sqrt{G_{k,q}^{\text{R}}( \boldsymbol{r}^{\text{R}} )}}{\partial \boldsymbol{r}^{\text{R}}}\boldsymbol{t}_{k,q}^{\text{RU}}( \boldsymbol{r}^{\text{R}} ) +\sqrt{G_{k,q}^{\text{R}}( \boldsymbol{r}^{\text{R}} )}\frac{\boldsymbol{t}_{k,q}^{\text{RU}}( \boldsymbol{r}^{\text{R}} )}{\partial \boldsymbol{r}^{\text{R}}} \bigg]},
\\
&\,\,\frac{\partial \boldsymbol{B}( \boldsymbol{r} )}{\partial \boldsymbol{r}^{\text{B}}}=\sum_{p=1}^P{\gamma_p}\bigg[ \frac{\partial \sqrt{G_{p}^{\text{B}}( \boldsymbol{r}^{\text{B}} )}}{\partial \boldsymbol{r}^{\text{B}}}\sqrt{G_{p}^{\text{R}}( \boldsymbol{r}^{\text{R}} )}\boldsymbol{t}_{p}^{\text{BR}}( \boldsymbol{r}^{\text{B}} ) ( \boldsymbol{t}_{p}^{\text{RB}}( \boldsymbol{r}^{\text{R}} ) ) ^H+\sqrt{G_{p}^{\text{B}}( \boldsymbol{r}^{\text{B}} ) G_{p}^{\text{R}}( \boldsymbol{r}^{\text{R}} )}\frac{\partial \boldsymbol{t}_{p}^{\text{BR}}( \boldsymbol{r}^{\text{B}} )}{\partial \boldsymbol{r}^{\text{B}}}( \boldsymbol{t}_{p}^{\text{RB}}( \boldsymbol{r}^{\text{R}} ) ) ^H \bigg],
\\
&\,\,\frac{\partial \boldsymbol{B}( \boldsymbol{r} )}{\partial \boldsymbol{r}^{\text{R}}}=\sum_{p=1}^P{\gamma_p}\bigg[ \sqrt{G_{p}^{\text{B}}( \boldsymbol{r}^{\text{B}} )}\frac{\partial \sqrt{G_{p}^{\text{R}}( \boldsymbol{r}^{\text{R}} )}}{\partial \boldsymbol{r}^{\text{R}}}\boldsymbol{t}_{p}^{\text{BR}}( \boldsymbol{r}^{\text{B}}( \boldsymbol{r}^{\text{B}} ) ) ( \boldsymbol{t}_{p}^{\text{RB}}( \boldsymbol{r}^{\text{R}} ) ) ^H+\sqrt{G_{p}^{\text{B}}( \boldsymbol{r}^{\text{B}} ) G_{p}^{\text{R}}( \boldsymbol{r}^{\text{R}} )}\boldsymbol{t}_{p}^{\text{BR}}( \boldsymbol{r}^{\text{B}} ) \Big( \frac{\partial \boldsymbol{t}_{p}^{\text{RB}}( \boldsymbol{r}^{\text{R}} )}{\partial \boldsymbol{r}^{\text{R}}} \Big) ^{H} \bigg],
\\
& \overline{\ \ \ \ \ \ \ \ \ \ \ \ \ \ \ \ \ \ \ \ \ \ \ \ \ \ \ \ \ \ \ \ \ \ \ \ \ \ \ \ \ \ \ \ \ \ \ \ \ \ \ \ \ \ \ \ \ \ \ \ \ \ \ \ \ \ \ \ \ \ \ \ \ \ \ \ \ \ \ \ \ \ \ \ \ \ \ \ \ \ \ \ \ \ \ \ \ \ \ \ \ \ \ \ \ \ \ \ \ \ \ \ \ \ \ \ \ \ \ \ \ \ \ \ \ \ \ \ \ \ \ \ \ \ \ \ \ \ \ \ \ \ \ \ \ \ \ } \nonumber
\end{align}
\end{subequations}
\vspace{-36pt}
\end{figure*}

According to the sensing beampattern \eqref{eq:transmit_beam_pattern} and the sensing channel model \eqref{eq:channel_BS_target}, for the beampattern derivative,
\begin{align}
\label{eq:dPa}
\frac{\partial \mathcal{P}_a}{\partial \boldsymbol{r}^{\text{A}}}
=2\Re\Big\{\Big(\frac{\partial \boldsymbol{f}_{\text{S},a}(\boldsymbol{r})}{\partial \boldsymbol{r}^{\text{A}}}\Big)^{H}
\boldsymbol{W}\boldsymbol{W}^H\boldsymbol{f}_{\text{S},a}(\boldsymbol{r})\Big\},
\end{align}
where $\frac{\partial \boldsymbol{f}_{\text{S},a}(\boldsymbol{r})}{\partial \boldsymbol{r}^{\text{A}}}$ is computed in \eqref{eq:gradient_PGD_fS}.
\begin{figure*}
\begin{subequations}
\label{eq:gradient_PGD_fS}
\begin{align}
&\frac{\partial \boldsymbol{f}_{\text{S},a}\left( \boldsymbol{r}^{\text{B}},\boldsymbol{r}^{\text{R}} \right)}{\partial \boldsymbol{r}^{\text{B}}}= \frac{\partial \left( \sqrt{G_{a}^{\text{B}}\left( \boldsymbol{r}^{\text{B}} \right)}\boldsymbol{t}^{\text{BT}}\left( \boldsymbol{r}^{\text{B}} \right) \right)}{\partial \boldsymbol{r}^{\text{B}}}+ \frac{\partial \boldsymbol{B}\left( \boldsymbol{r}^{\text{B}},\boldsymbol{r}^{\text{R}} \right)}{\partial \boldsymbol{r}^{\text{B}}}\mathrm{diag}\left( \boldsymbol{\theta } \right) \sqrt{G_{a}^{\text{R}}\left( \boldsymbol{r}^{\text{R}} \right)}\,\boldsymbol{t}^{\text{RT}}\left( \boldsymbol{r}^{\text{R}} \right) ,
\\
&\frac{\partial \boldsymbol{f}_{\text{S},a}\left( \boldsymbol{r}^{\text{B}},\boldsymbol{r}^{\text{R}} \right)}{\partial \boldsymbol{r}^{\text{R}}}= \frac{\partial \boldsymbol{B}\left( \boldsymbol{r}^{\text{B}},\boldsymbol{r}^{\text{R}} \right)}{\partial \boldsymbol{r}^{\text{R}}}\mathrm{diag}\left( \boldsymbol{\theta } \right) \sqrt{G_{a}^{\text{R}}}\boldsymbol{t}^{\text{RT}}+ \boldsymbol{B}\left( \boldsymbol{r}^{\text{B}},\boldsymbol{r}^{\text{R}} \right) \mathrm{diag}\left( \boldsymbol{\theta } \right) \frac{\partial \left( \sqrt{G_{a}^{\text{R}}\left( \boldsymbol{r}^{\text{R}} \right)}\,\boldsymbol{t}^{\text{RT}}\left( \boldsymbol{r}^{\text{R}} \right) \right)}{\partial \boldsymbol{r}^{\text{R}}}
\\
& \overline{\ \ \ \ \ \ \ \ \ \ \ \ \ \ \ \ \ \ \ \ \ \ \ \ \ \ \ \ \ \ \ \ \ \ \ \ \ \ \ \ \ \ \ \ \ \ \ \ \ \ \ \ \ \ \ \ \ \ \ \ \ \ \ \ \ \ \ \ \ \ \ \ \ \ \ \ \ \ \ \ \ \ \ \ \ \ \ \ \ \ \ \ \ \ \ \ \ \ \ \ \ \ \ \ \ \ \ \ \ \ \ \ \ \ \ \ \ \ \ \ \ \ \ \ \ \ \ \ \ \ \ \ \ \ \ \ \ \ \ \ \ \ \ \ \ \ \ } \nonumber
\end{align}
\end{subequations}
\vspace{-36pt}
\end{figure*}
It is also convenient to gather partials into Jacobians
\(\boldsymbol{J}_{\text{S},a}^{(\text{A})} \triangleq \left[\frac{\partial \boldsymbol{f}_{\text{S},a}}{\partial r_{\mathrm{x}}^{\text{A}}},\frac{\partial \boldsymbol{f}_{\text{S},a}}{\partial r_{\mathrm{y}}^{\text{A}}},\frac{\partial \boldsymbol{f}_{\text{S},a}}{\partial r_{\mathrm{z}}^{\text{A}}}\right]\in\mathbb{C}^{M\times 3}\) similar to \eqref{eq:gradient_PGD_t},
so that \eqref{eq:dPa} compacts to
\begin{align}
\label{eq:dPa-compact}
\frac{\partial \mathcal{P}_a}{\partial \boldsymbol{r}^{\text{A}}}
=2\,\Re\!\Big\{\boldsymbol{J}_{\text{S},a}^{(\text{A})\,H}\boldsymbol{W}\boldsymbol{W}^H\boldsymbol{f}_{\text{S},a}\Big\}\in\mathbb{R}^3.
\end{align}

\subsubsection{Projected-gradient framework for the rotation block.}

Let $\mathcal{B}$ be the box of array rotation angle limits in (P4)
\begin{align}
\mathcal{B}=\{\boldsymbol{r}:\boldsymbol{r}_{\text{lb}}^{}\preceq \boldsymbol{r}\preceq \boldsymbol{r}_{\text{ub}}^{}\},
\end{align}
where we denote $\boldsymbol{r}_{\text{lb}}\triangleq[ ( \boldsymbol{r}_{\text{lb}}^{\text{B}} ) ^T,( \boldsymbol{r}_{\text{lb}}^{\text{R}} ) ^T ] ^T
$ and $\boldsymbol{r}_{\text{ub}}\triangleq [ ( \boldsymbol{r}_{\text{ub}}^{\text{B}} ) ^T,( \boldsymbol{r}_{\text{ub}}^{\text{R}} ) ^T ] ^T$, and here we let $\boldsymbol{r}_{\text{lb}}^{\text{A}}=[ r_{\mathrm{x},\min}^{\text{A}},r_{\mathrm{y},\min}^{\text{A}},r_{\mathrm{z},\min}^{\text{A}}]$ and $\boldsymbol{r}_{\text{ub}}^{\text{A}}=[ r_{\mathrm{x},\max}^{\text{A}},r_{\mathrm{y},\max}^{\text{A}},r_{\mathrm{z},\max}^{\text{A}}]$ corresponding to \eqref{p1:Bbox} and $\eqref{p1:Rbox}$, respectively, for brevity. 
Let $\Pi_{\mathcal{B}}(\cdot)$ be the Euclidean projection onto $\mathcal{B}$.
We perform projected gradient ascent of $F\left( \boldsymbol{r} \right)$ on the feasible region with the projected step 
\begin{align}
\label{eq:PGD_projection}
\boldsymbol{\beta}\left( \alpha \right) =\Pi _{\mathcal{B}}\left( \boldsymbol{r}+\alpha \nabla _{\boldsymbol{r}}F\left( \boldsymbol{r} \right) \right) -\boldsymbol{r},
\end{align}
for any step size $\alpha>0$. Define the projected-gradient mapping $\boldsymbol{\upsilon}_{\alpha}\left( \boldsymbol{r} \right) \triangleq \frac{1}{\alpha}\boldsymbol{\beta}\left( \alpha \right)$, for any $\alpha>0$, $\boldsymbol r^\star$ is a first-order stationary point of (P4) if and only if $\boldsymbol{\upsilon}_{\alpha}\left( \boldsymbol{r}^{\star} \right) =\boldsymbol{0}$, according to the KKT condition. Hence $\|\boldsymbol{\upsilon}_{\alpha}(\boldsymbol r)\|_2$ is a valid criticality measure for stopping. 
Because $\mathcal{B}$ is a box, the projection \eqref{eq:PGD_projection} reduces to the component-wise clipping
\begin{align}
\label{eq:PG-box-component}
 \boldsymbol{\beta}^{(\ell)}(\alpha) =\mathrm{clip}( \boldsymbol{r}^{\left( \ell \right)}+\alpha \nabla_{\boldsymbol{r}} F^{( \ell )}( \boldsymbol{r})) _{[ r_{\text{lb}},r_{\text{ub}}]} -\boldsymbol{r}^{(\ell)},
\end{align}
where the clip operation is defined by $\mathrm{clip}\left( \mathfrak{a} \right) _{\left[ \mathfrak{b} ,\mathfrak{c} \right]}=\min \left\{ \max \left\{ \mathfrak{a},\mathfrak{b} \right\} ,\mathfrak{c} \right\}$ that acts element-wise. Given the previous $\left( \boldsymbol{r},F\left( \boldsymbol{r} \right) ,\nabla F\left( \boldsymbol{r} \right) \right)^{(\ell-1)}$, the step size uses the Barzilai–Borwein initialization
\begin{align}
\label{eq:BB1}
\alpha _{\text{BB}}=\frac{\|\boldsymbol{r}^{\left( \ell \right)}-\boldsymbol{r}^{\left( \ell -1 \right)}\|^2}{\left( \boldsymbol{r}^{\left( \ell \right)}-\boldsymbol{r}^{\left( \ell -1 \right)} \right) ^T\left( \nabla F\left( \boldsymbol{r}^{\left( \ell \right)} \right) -\nabla F\left( \boldsymbol{r}^{\left( \ell -1 \right)} \right) \right)},
\end{align}
and clip to $[\alpha_{\min},\alpha_{\max}]$ to ensure robustness. An Armijo backtracking is then performed until the ascent rule holds
\begin{align}
\label{eq:PGD_backtracking}
F\left( \boldsymbol{r}+\boldsymbol{\beta}\left( \alpha \right) \right) \ge F\left( \boldsymbol{r} \right) +\kappa_1\nabla _{\boldsymbol{r}}F\left( \boldsymbol{r} \right) ^T\boldsymbol{\beta}\left( \alpha \right),
\end{align}
with $\kappa_1\!\in\!(0,1)$ and the backtracking using $\alpha\!\leftarrow\!\kappa_2\alpha$, $\kappa_2\in(0,1)$. The iteration terminates when it meets the stopping criteria that the projected gradient norm is smaller than a threshold $\varepsilon_{\text{pg}}$, i.e., $\|\boldsymbol{\upsilon}_{\alpha}(\boldsymbol r^{(\ell)})\|_2\ \le\ \varepsilon_{\text{pg}}$, or the relative step is smaller than a threshold $\varepsilon_{\text{rel}}$, i.e., $\frac{\|\boldsymbol r^{(\ell)}-\boldsymbol r^{(\ell-1)}\|_2}{\max\{1,\|\boldsymbol r^{(\ell-1)}\|_2\}}\ \le\ \varepsilon_{\text{rel}}$. 
For box constraints, $\boldsymbol{\upsilon}_{\alpha}(\boldsymbol r^{(\ell)})$ admits the following component-wise form for $\imath$-th entry, according to \eqref{eq:stacked_rotation}, independent of $\alpha$
\begin{align}
\label{eq:PG-box}
\!\!\!\big[\boldsymbol{\upsilon}_{\alpha}(\boldsymbol r^{(\ell)})\big]_{\imath}
\!=\!\!
\begin{cases}
[\nabla _{\boldsymbol{r}}F^{\left( \ell \right)}( \boldsymbol{r} )]_{\imath}
, \!& \!\!\!r_{\text{lb},i} \! \!< \!r_i \!< \!r_{\text{ub},i}, 
\\
\max\{0, [\nabla _{\boldsymbol{r}}F^{( \ell)}( \boldsymbol{r})]_{\imath}\},\! & \!\!\!r_{\imath} =r_{\text{lb},\imath},
\\
\min\{0, [\nabla _{\boldsymbol{r}}F^{( \ell)}( \boldsymbol{r})]_{\imath}\}, \!& \! \!\!r_{\imath} =r_{\text{ub},\imath}.
\end{cases} \!\!\!\!
\end{align}
The gradient cost is dominated by Jacobians of rotation-related steering vectors and derivatives with matrix-chain products. Because the directional-gain model is piecewise-smooth, we use subgradients in \eqref{eq:PG-box} along the visibility boundary to ensure stable line search. 

\begin{algorithm}[t]
\caption{PGD for Array Rotation Subproblem (P4)}
\label{alg:PGD-rot}
\small
\begin{algorithmic}[1]
\STATE \textbf{Input:} fixed $(\boldsymbol W,\boldsymbol\theta)$, box $\mathcal{B}$, tolerances $\varepsilon_{\mathrm{pg}},\varepsilon_{\mathrm{rel}}>0$, and maximum iterations $I_{\text{PGD}}^{\text{max}}$.
\STATE \textbf{Initialization:} initial point $\boldsymbol r^{(0)} \in \mathcal{B}$ 
\FOR{$\ell = 0, 1,\ldots, I_{\text{PGD}}^{\text{max}}-1$}
  \STATE Evaluate the cost $F(\boldsymbol r^{(\ell)})$ and the gradient $\nabla_{\boldsymbol r}F(\boldsymbol r^{(\ell)})$ using \eqref{p4:objective}-\eqref{eq:dPa-compact}.
  \IF{$\ell>0$}
    \STATE Barzilai-Borwein step size initialization $\alpha^{(\ell)}\!\leftarrow\!\alpha_{\text{BB}}$ via \eqref{eq:BB1}.
  \ELSE 
    \STATE $\alpha^{(\ell)} \leftarrow \alpha_0$.
  \ENDIF
  \STATE Find step size $\alpha^{(\ell)}$ via Armijo backtracking to satisfy \eqref{eq:PGD_backtracking}.
  \STATE Determine projected step $\boldsymbol{\beta}^{(\ell)}(\alpha)$ via \eqref{eq:PG-box-component}
  \STATE Update optimization variables by $\boldsymbol r^{(\ell+1)} =\boldsymbol r^{(\ell)} + \boldsymbol{\beta}^{(\ell)}(\alpha)$.
  \STATE Compute projected gradient mapping $\boldsymbol{\upsilon}_{\alpha}(\boldsymbol r^{(\ell)})$ via \eqref{eq:PG-box}.
  \IF{$\|\boldsymbol{\upsilon}_{\alpha}(\boldsymbol r^{(\ell+1)})\|_2 \le \varepsilon_{\mathrm{pg}}$ \OR
       $\dfrac{\|\boldsymbol r^{(\ell+1)} - \boldsymbol r^{(\ell)}\|_2}
              {\max\{1,\|\boldsymbol r^{(\ell)}\|_2\}} \le \varepsilon_{\mathrm{rel}}$}
    \STATE \textbf{break};
  \ENDIF
\ENDFOR
\STATE \textbf{Output:} $\boldsymbol r^{\text{B}} \leftarrow \boldsymbol r^{(\ell+1)}_{1:3}$,
$\boldsymbol r^{\text{R}} \leftarrow \boldsymbol r^{(\ell+1)}_{4:6}$.
\end{algorithmic}
\end{algorithm}

\subsubsection{Overall algorithm and complexity analysis for subproblem 3}
Algorithm 3 summarizes the overall projected gradient method. For the per-iteration computational complexity, directly forming $\boldsymbol B$ and the Jacobians as dense $M\times N$ matrices yields $\mathcal O(PMN)$ computations per rotation axis, and evaluating the sensing gradients over $A$ grids would add $\mathcal O(AMN)$. However, using the rank-$P$ factorization for $\boldsymbol B$ and its derivatives, all costly matrix–vector products, involving $\boldsymbol B$ and $\partial\boldsymbol B/\partial r_{\imath}^{\text{A}}$, are computed without materializing $M\times N$ blocks and costing $\mathcal O(NP)$ plus $\mathcal O(MP)$.
With these reuses, one PGD iteration with full gradient of $F$ has the leading cost: i) precomputation on steering vectors and their Jacobians with $\mathcal{O}\big(P(M+N)\big)$, ii) sensing and communication channel–Jacobian products $\mathcal O\big((A+K)P(N+M)\big)$, iii) use $\boldsymbol W(\boldsymbol W^H\boldsymbol f)$ to apply $\boldsymbol W\boldsymbol W^H$ in \eqref{eq:gradient_PGD_C} and \eqref{eq:dPa-compact} with  $\mathcal O\!\big((M+K)M(A+K)\big)$. Therefore, the per-iteration complexity scales as $\mathcal{O}\big(P(M+N)(A+K+1)+(M+K)M(A+K)\big)$ with a constant coefficient affected by backtracking searches.

\vspace{-12pt}
\subsection{Overall algorithm for the original problem}
Finally, Algorithm~\ref{alg:ao_nmse} summarizes the overall AO procedure for (P1). At the $\zeta$-th outer iteration, we first update the NMSE scaling parameter by $\iota^{(\zeta)}\leftarrow \iota^\star$ in \eqref{eq:iota_star}. Then, with the other blocks fixed, we update $\boldsymbol W$ by solving (P2), update $\boldsymbol\theta$ by solving (P3) on the complex circle manifold, and update $(\boldsymbol r^{\text{B}},\boldsymbol r^{\text{R}})$ by solving (P4) via projected gradient ascent under box constraints.
Since each block update is designed to be non-decreasing in the objective of (P1) with $\iota$ fixed, and the $\iota$-update minimizes $\text{NMSE}(\iota)$ for fixed $(\boldsymbol W,\boldsymbol\theta,\boldsymbol r^{\text{B}},\boldsymbol r^{\text{R}})$, the overall objective in \eqref{p1:objective} is monotonically non-decreasing over outer iterations. The algorithm thus converges to a stationary point of (P1) under standard regularity assumptions. However, due to the intrinsic non-convexity of (P1), global optimality is generally intractable, and the proposed AO-based algorithm obtains a suboptimal solution for feasible system design.

\begin{algorithm}[t]
\caption{AO-based Algorithm for (P1)}
\label{alg:ao_nmse}
\begin{algorithmic}[1]
\STATE \textbf{Input:} channel and angular parameters, $P_{\text{B}}$, $\rho$, desired pattern $\{\mathcal{P}_{\text{d},a}\}_{a=1}^{A}$, tolerances $\epsilon_{\mathrm{AO}}$.
\STATE \textbf{Initialize:} feasible $\boldsymbol W^{(0)}$, $\boldsymbol\theta^{(0)}$, and $(\boldsymbol r^{\text{B}},\boldsymbol r^{\text{R}})^{(0)}$ and set $\zeta=0$.
\REPEAT
\STATE Compute current beampattern samples $\boldsymbol{p}^{(\zeta)}=[\mathcal{P}_1,\ldots,\mathcal{P}_A]^{T}$ and update $\iota^{(\zeta)} \leftarrow \iota^\star$ via \eqref{eq:iota_star}.
\STATE Update $\boldsymbol W^{(\zeta+1)}$ by solving (P2) via QT, MM, and bisection search.
\STATE Update $\boldsymbol\theta^{(\zeta+1)}$ by solving (P3) via RCG on the complex circle manifold.
\STATE Update $(\boldsymbol r^{\text{B}},\boldsymbol r^{\text{R}})^{(\zeta+1)}$ by solving (P4) via projected gradient ascent with Armijo backtracking.
\STATE $\zeta\leftarrow \zeta+1$.
\UNTIL{the relative increase of the objective in (P1) is below $\epsilon_{\mathrm{AO}}$ (or a maximum iteration is reached).}
\STATE \textbf{Output:} $(\boldsymbol W,\boldsymbol\theta,\boldsymbol r^{\text{B}},\boldsymbol r^{\text{R}})$.
\end{algorithmic}
\end{algorithm}

\section{Simulation Results}

This section evaluates the proposed low-altitude ISAC design with cooperatively rotatable BS and RIS arrays. For performance characterization, we report three metrics: the downlink sum rate $R\triangleq \sum_{k=1}^{K}\log_2(1+\mathrm{SINR}_k)$,
the beampattern mismatch quantified by the NMSE in \eqref{eq:nmse_def} with $\iota$ updated by \eqref{eq:iota_star} at each AO outer iteration,
and the weighted ISAC utility $\mathcal{U}\triangleq \sum_{k=1}^{K}\log_2(1+\mathrm{SINR}_k)-\rho\text{NMSE}$ of (P1). 

\subsection{Simulation Setup and Benchmarks}

Unless otherwise stated, we set $\rho=10$. The BS located at $(0,0,0)$employs a $M=M_{\text{col}}M_{\text{row}}=2\times 2$ UPA with inter-antenna spacing $\lambda/2$ at carrier wavelength $\lambda=0.1$~m.
The RIS is a $6\times 6$ UPA with element spacing $\lambda/2$, located at $(10,0,0)$~m relative to the BS.
The small-scale fading of all links follows a rotation-aware geometry-based multipath model: each BS-user and RIS-user channel contains $L_k=Q_k=2$ paths, and the BS-RIS channel contains $P=2$ paths.
The azimuth and elevation angles are independently generated with $\varphi\in[-\pi,\pi]$ and $\vartheta\in[-\pi/3,\pi/3]$, and the complex path gains are drawn from independent circularly symmetric Gaussian distributions with normalized variance.
For sensing, we discretize the angular region $\phi\in[-\pi,\pi]$ and $\vartheta\in[-\pi/4,\pi/4]$ with $A_{\mathrm{az}}=11$ azimuth sampling points and $A_{\mathrm{el}}=6$ elevation points, resulting in $A=A_{\mathrm{az}}A_{\mathrm{el}}=66$ grid points.
The desired pattern is a spotlight-type template consisting of two angular sectors, and the NMSE is computed with the optimal scaling $\iota^\star$ in \eqref{eq:iota_star}. The maximum rotation angle of the active and passive array is set by $\pi/2$ for each Euler angle in \eqref{p1:Bbox} and \eqref{p1:Rbox} by $r_{\text{i},\min}^{\text{R}}=r_{\text{i},\min}^{\text{B}}=-\frac{\pi}{2},r_{\text{i},\max}^{\text{R}}=r_{\text{i},\max}^{\text{B}}=\frac{\pi}{2},\text{i}\in \{\mathrm{x},\mathrm{y},\mathrm{z}\}$. 
We consider two settings on the directional radiation pattern gain of the BS antenna and the RIS element:
\emph{half-space isotropic} ($b^{\text{B}}=b^{\text{R}}=0$) and \emph{directional} ($b^{\text{B}}=b^{\text{R}}=2$), where the latter yields a narrower mainlobe and stronger gain variations over directions.

We compare the proposed jointly rotatable BS and RIS scheme (``Rot-BS \& Rot-RIS") with the following baselines:
(i) ``Rot-BS \& Fix-RIS", where only the BS array is rotatable while the RIS array orientation is fixed, 
(ii) ``Fix-BS \& Rot-RIS", where only the RIS array is rotatable while the BS array orientation is fixed, 
(iii) ``Fix-BS \& Fix-RIS", where both BS and RIS arrays are fixed,
(iv) ``Rot-BS \& No-RIS", where the RIS–user link is disabled by setting the RIS–user path gains to zero, and (v) ``Fix-BS \& No-RIS".
All reported curves are obtained by averaging the communication and sensing metrics over 100 independent channel realizations.

\setlength{\abovecaptionskip}{0pt}
\begin{figure}[t]
\centering
\includegraphics[width=2.9in]{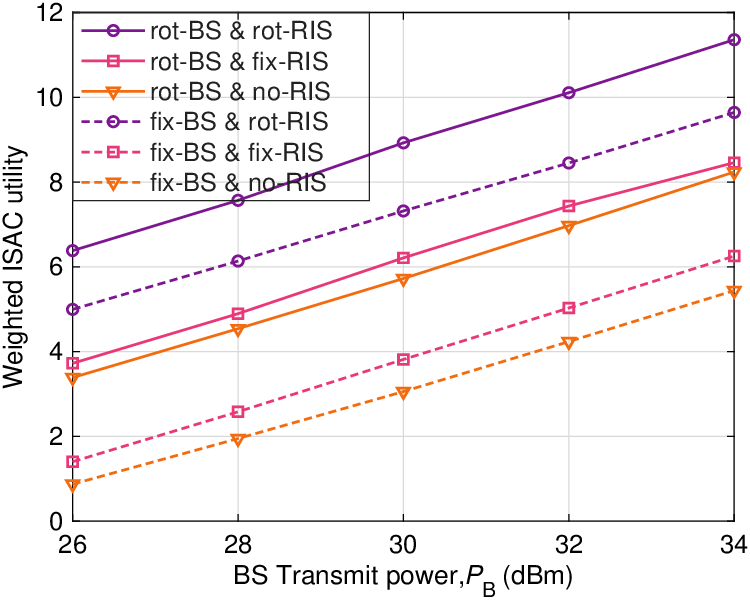}
\caption{Weighted ISAC utility versus maximum transmit power under $b^{\text{R}}=b^{\text{B}}=0$ with $\rho=10$, $N=6\times6$, $M=2\times2$ and $K=2$.} 
\end{figure}
\begin{figure}[t]
\centering
\includegraphics[width=2.9in]{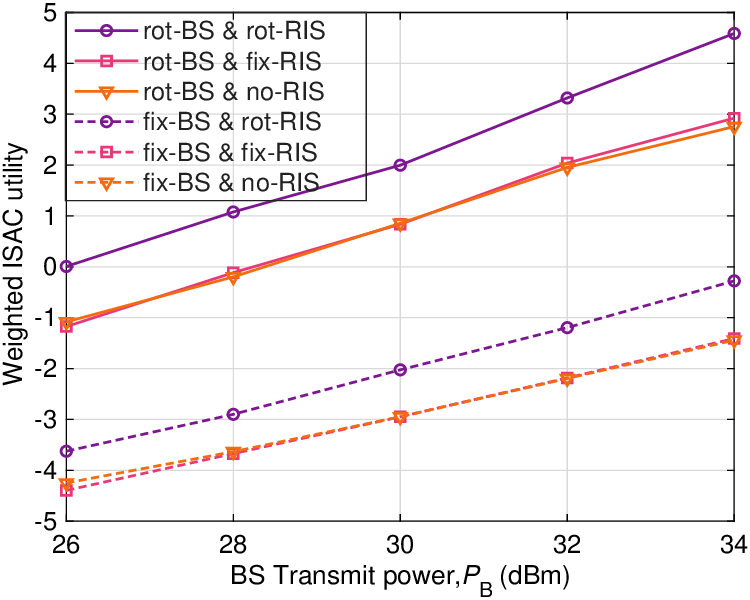}
\caption{Weighted ISAC utility versus maximum transmit power under $b^{\text{R}}=b^{\text{B}}=2$ with $\rho=10$, $N=6\times6$, $M=2\times2$ and $K=2$.} 
\vspace{-12pt}
\end{figure}

\begin{figure}[t]
\centering
\includegraphics[width=2.9in]{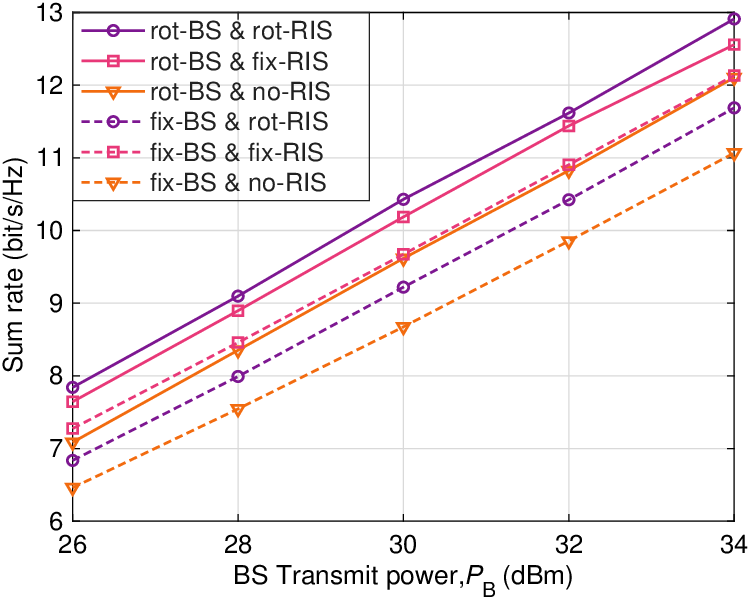}
\caption{Sum rate versus maximum transmit power under $b^{\text{R}}=b^{\text{B}}=0$ with $\rho=10$, $N=6\times6$, $M=2\times2$ and $K=2$.} 
\end{figure}
\begin{figure}[t]
\centering
\includegraphics[width=2.9in]{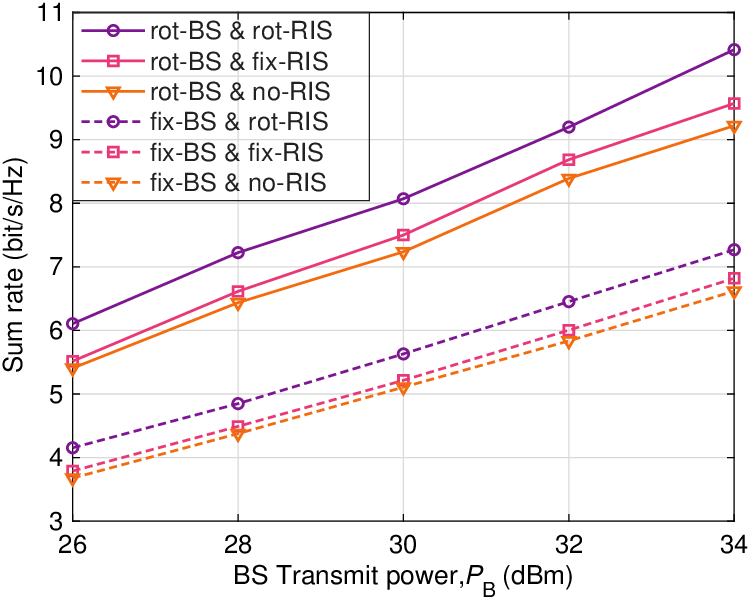}
\caption{Sum rate versus maximum transmit power under $b^{\text{R}}=b^{\text{B}}=2$ with $\rho=10$, $N=6\times6$, $M=2\times2$ and $K=2$.} 
\end{figure}

\begin{figure}[t]
\centering
\includegraphics[width=2.9in]{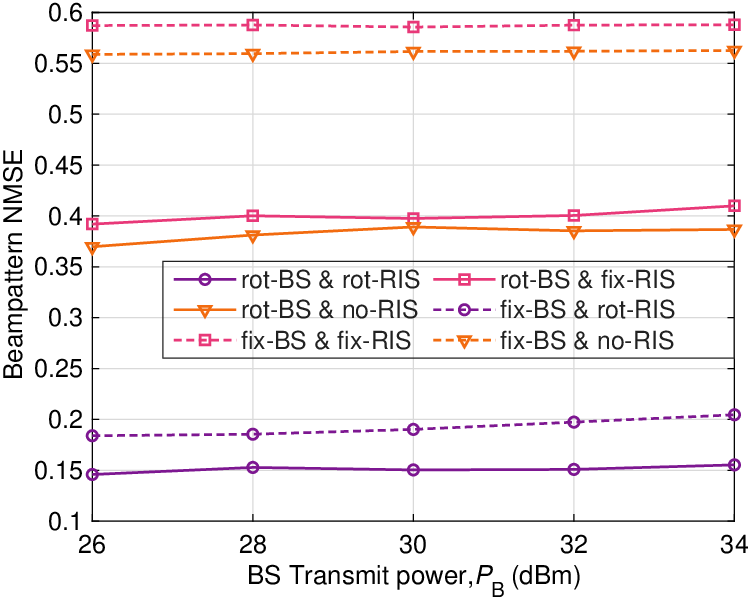}
\caption{Beampattern NMSE versus maximum transmit power under $b^{\text{R}}=b^{\text{B}}=0$ with $\rho=10$, $N=6\times6$, $M=2\times2$ and $K=2$.} 
\end{figure}
\begin{figure}[t]
\centering
\includegraphics[width=2.9in]{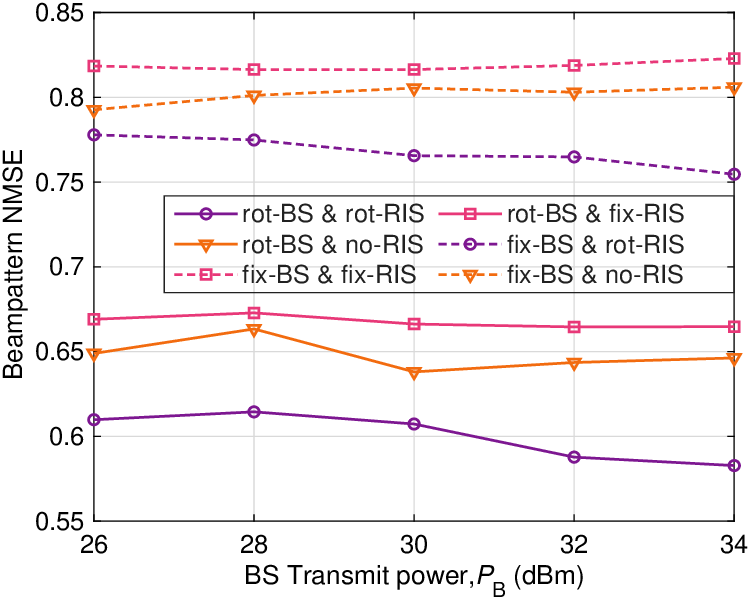}
\caption{Beampattern NMSE versus maximum transmit power under $b^{\text{R}}=b^{\text{B}}=2$ with $\rho=10$, $N=6\times6$, $M=2\times2$ and $K=2$.} 
\vspace{-12pt}
\end{figure}

\vspace{-12pt}
\subsection{Impact of Transmit Power}

Figs.~2-7 evaluate the impact of the BS transmit-power budget $P_{\text{B}}$ under $M=2\times 2$, $N=6\times 6$, and $K=2$. Figs.~2-3 plot the weighted utility $\mathcal U$, while Figs.~4-5 and Figs.~6-7 plot the associated sum rate and NMSE, respectively.

\emph{Half-space isotropic elements ($b=0$):}
From Figs.~2, 4, and 6, increasing $P_{\text{B}}$ monotonically improves the achievable utility and sum rate for all schemes, as higher power boosts the effective SNR.
Meanwhile, the NMSE changes mildly with $P_{\text{B}}$, indicating that with the optimal scaling $\iota^\star$, the NMSE is dominated by spatial shape mismatch rather than global amplitude. Across the entire power range, ``Rot-BS \& Rot-RIS" achieves the best performance in both $\mathcal U$ and $R$, together with the smallest NMSE. This gain comes from the additional geometric DoF provided by jointly rotating the active and passive arrays, which reshapes the array manifolds and effectively enlarges the feasible set of beampatterns under the same electronic beamforming resources. Moreover, the RIS-assisted schemes significantly outperform the no-RIS baselines, confirming that the additional passive aperture not only enhances the downlink channels but also provides additional spatial controllability for sensing beampattern shaping.

\emph{Directional elements ($b=2$):}
Figs.~3, 5, and 7 show that the benefit of joint rotation becomes substantially more pronounced when practical directional element patterns are employed. In this case, misalignment between the array boresight and the desired propagation directions can severely reduce the effective channel gain (and thus increase NMSE), potentially driving the weighted utility to negative values for fixed-orientation baselines. By contrast, ``Rot-BS \& Rot-RIS" maintains the highest utility and sum rate and achieves the lowest NMSE, since it can simultaneously align the BS and RIS boresights with the dominant user/target directions and avoid low-gain regions. These results highlight a key LAWN-specific insight: mechanical geometry control is most valuable when the system operates with directional arrays and/or becomes interference-limited, where purely electronic beamforming alone is insufficient to reliably meet both communication and sensing objectives.

\begin{figure}[t]
\centering
\includegraphics[width=2.9in]{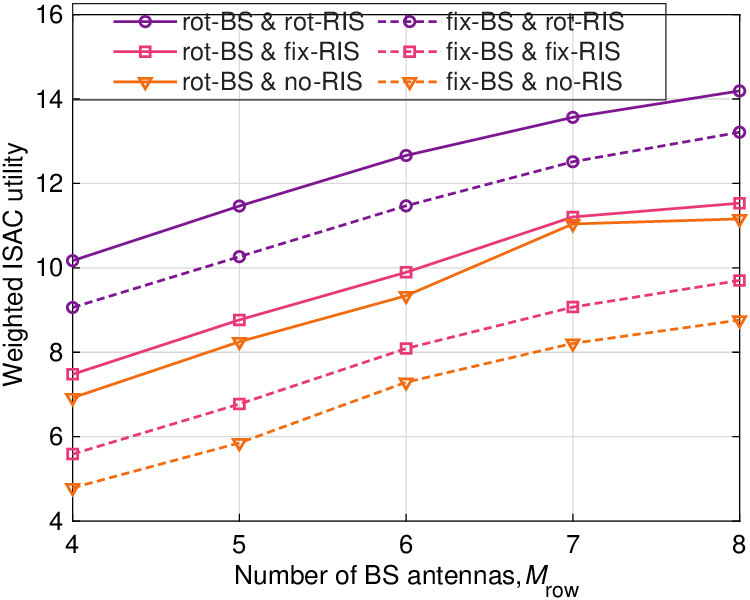}
\caption{Weighted ISAC utility  the number of BS antennas under $b^{\text{R}}=b^{\text{B}}=0$ with $\rho=10$, $N=6\times6$, $M_{\text{col}}=1$, $P=30$ dBm, and $K=3$.} 
\end{figure}
\begin{figure}[t]
\centering
\includegraphics[width=2.9in]{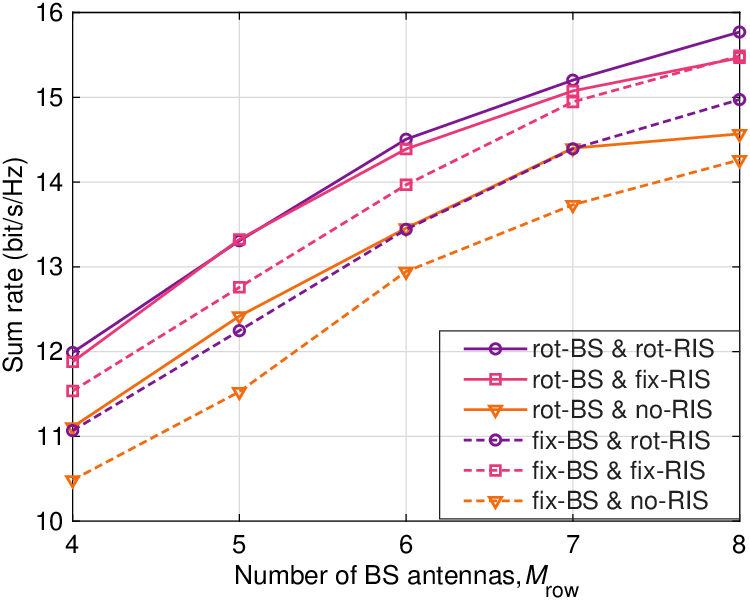}
\caption{Sum rate versus the number of BS antennas under $b^{\text{R}}=b^{\text{B}}=0$ with $\rho=10$, $N=6\times6$, $M_{\text{col}}=1$, $P=30$ dBm, and $K=3$.} 
\vspace{-12pt}
\end{figure}
\begin{figure}[t]
\centering
\includegraphics[width=2.9in]{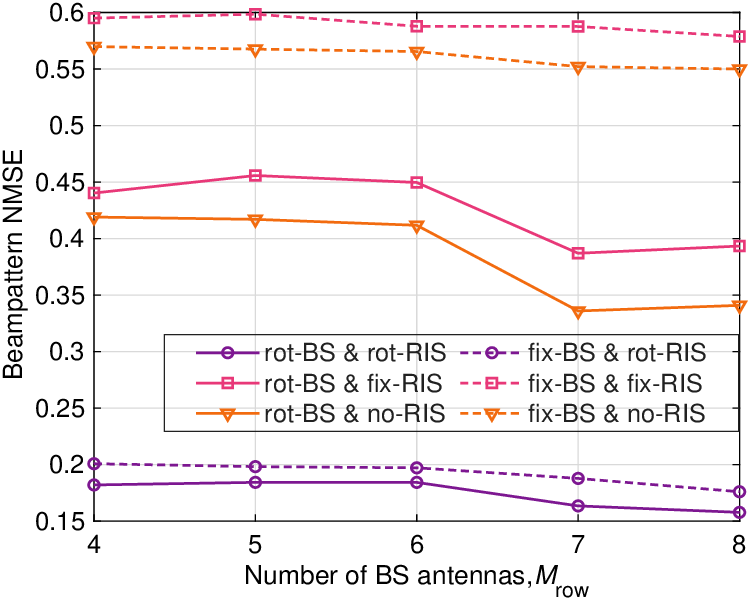}
\caption{Beampattern NMSE versus the number of BS antennas under $b^{\text{R}}=b^{\text{B}}=0$ with $\rho=10$, $N=6\times6$, $M_{\text{col}}=1$, $P=30$ dBm, and $K=3$.} 
\end{figure}

\subsection{Impact of BS Array Size}

Figs.~8-10 study the impact of the BS array size by increasing $M_{\text{row}}$ for a $1\times M_{\text{row}}$ BS array under $K=3$ and $P_{\text{B}}=30$~dBm (with $b=0$). As $M_{\text{row}}$ increases, the weighted utility and sum rate (Figs.~8-9) increase steadily because a larger aperture provides higher array gain and more spatial DoF to separate users while shaping the sensing beampattern. In addition, the NMSE decreases with $M_{\text{row}}$ (Fig.~10), showing that more antennas enable a finer beampattern approximation for the same desired template. Notably, the proposed ``Rot-BS \& Rot-RIS" scheme consistently achieves the best utility-rate-NMSE trade-off across all $M_{\text{row}}$, while the partially rotatable and fully fixed schemes exhibit higher NMSE floors.
This indicates that array scaling and geometry adaptation are complementary: increasing $M$ expands the electronic beamforming space, while rotation expands the \emph{geometric} manifold of feasible array responses, and the combination yields the most robust ISAC performance.

\begin{figure}[t]
\centering
\includegraphics[width=2.9in]{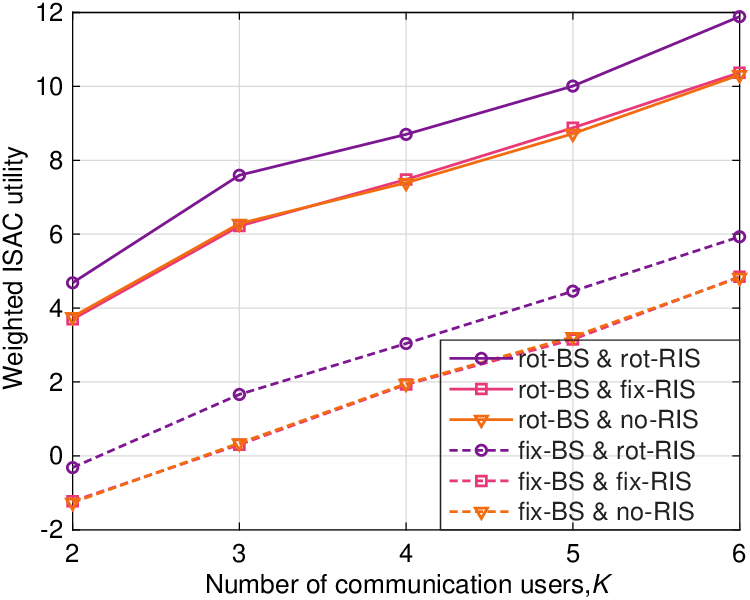}
\caption{Weighted ISAC utility versus the number of communication users under $b^{\text{R}}=b^{\text{B}}=2$ with $\rho=10$, $N=6\times6$, $M=2\times4$, and $P=30$ dBm.} 
\vspace{-12pt}
\end{figure}
\begin{figure}[t]
\centering
\includegraphics[width=2.9in]{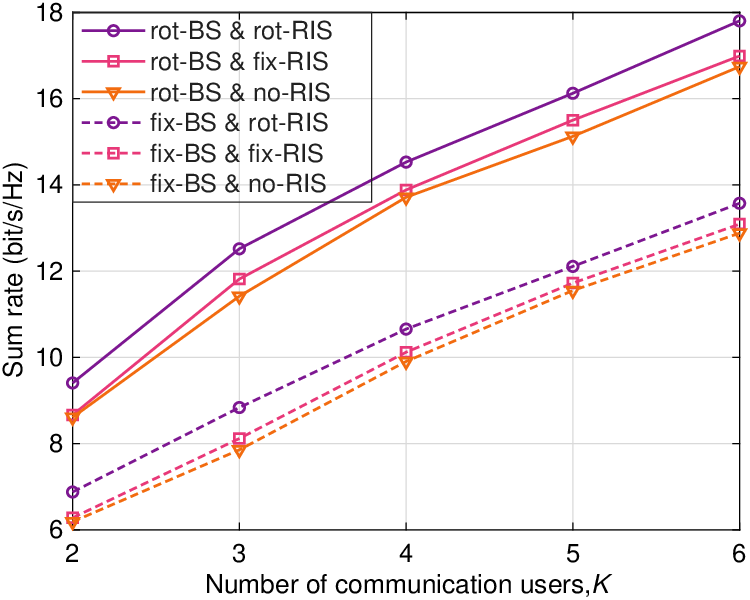}
\caption{Sum rate versus the number of communication users under $b^{\text{R}}=b^{\text{B}}=2$ with $\rho=10$, $N=6\times6$, $M=2\times4$, and $P=30$ dBm.} 
\end{figure}
\begin{figure}[t]
\centering
\includegraphics[width=2.9in]{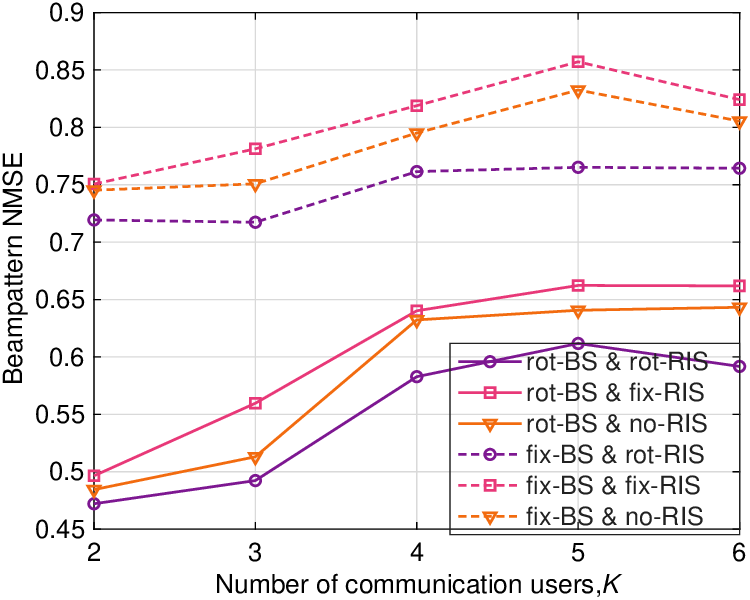}
\caption{Beampattern NMSE versus the number of communication users under $b^{\text{R}}=b^{\text{B}}=2$ with $\rho=10$, $N=6\times6$, $M=2\times4$, and $P=30$ dBm.} 
\vspace{-12pt}
\end{figure}

\subsection{Impact of User Loading}

Figs.~11-13 investigate user scaling under a $2\times 4$ BS array, $N=6\times 6$, $P_{\text{B}}=30$~dBm, and directional elements ($b=2$). As the number of users $K$ increases, the sum rate increases due to multiuser spatial multiplexing gain (Fig.~12), and consequently the weighted utility also increases (Fig.~11). However, Fig.~13 shows that the NMSE generally increases with $K$, which reflects the intrinsic ISAC tension: serving more users requires allocating more spatial resources to interference management and data transmission, leaving less flexibility for beampattern shaping. Importantly, ``Rot-BS \& Rot-RIS" consistently yields the highest utility and sum rate while achieving the lowest NMSE across all $K$. This demonstrates that joint BS and RIS rotation is particularly beneficial in heavily loaded regimes, where geometric adaptation provides additional leverage to mitigate interference and preserve sensing beampattern fidelity under directional element constraints.

\begin{figure}[t]
\centering
\includegraphics[width=2.9in]{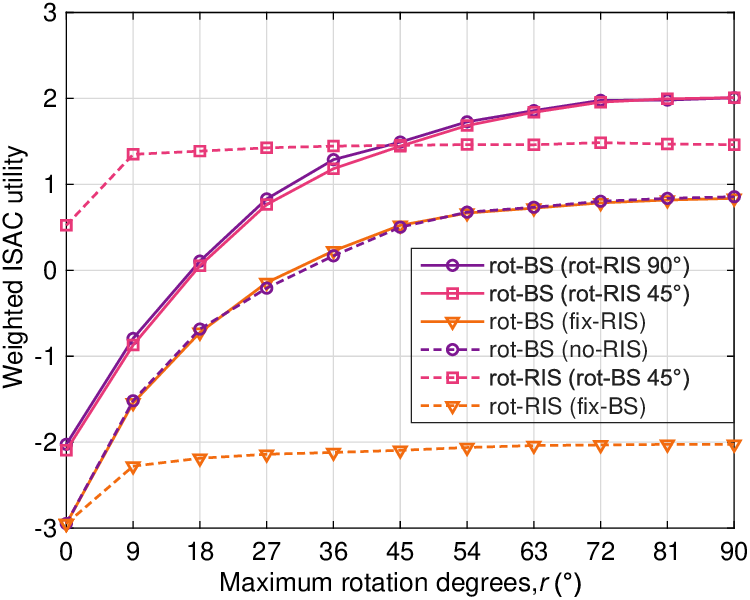}
\caption{Weighted ISAC utility versus the maximum rotation range $r$ (degrees) under $b^{\text{R}}=b^{\text{B}}=2$ with $N=6\times6$, $M=2\times2$, and $P_{\text{B}}=30$ dBm.}
\end{figure}
\begin{figure}[t]
\centering
\includegraphics[width=2.9in]{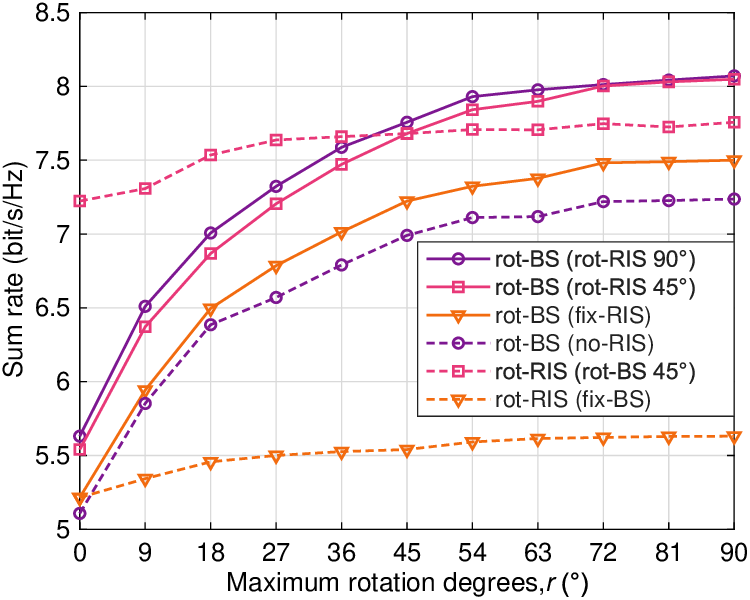}
\caption{Sum rate versus the maximum rotation range $r$ under $b^{\text{R}}=b^{\text{B}}=2$ with $N=6\times6$, $M=2\times2$, and $P_{\text{B}}=30$ dBm.}
\vspace{-12pt}
\end{figure}
\begin{figure}[t]
\centering
\includegraphics[width=2.9in]{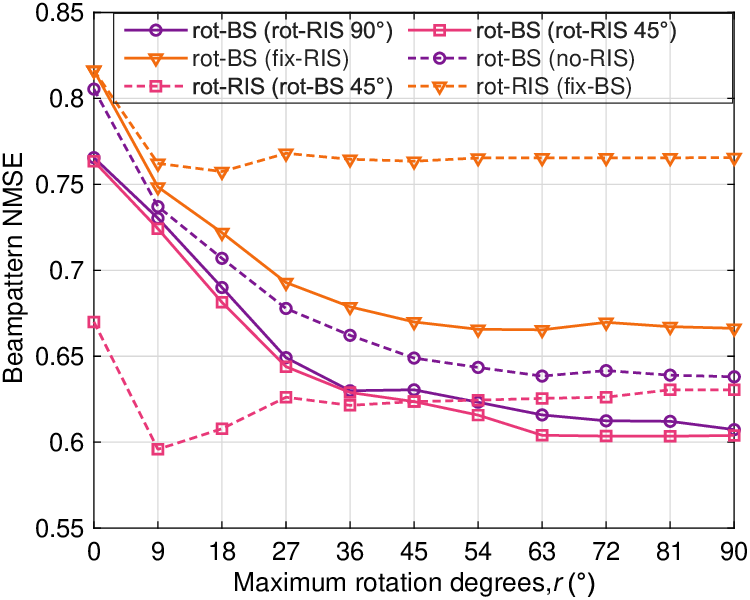}
\caption{Beampattern NMSE versus the maximum rotation range $r$ (degrees) under $b^{\text{R}}=b^{\text{B}}=2$ with $N=6\times6$, $M=2\times2$, and $P_{\text{B}}=30$ dBm.}
\end{figure}

\subsection{Impact of Rotation Range Constraints}

Figs.~14-16 evaluate the effect of mechanical rotation limits under $b=2$ with $M=2\times 2$, $N=6\times 6$, and $P_{\text{B}}=30$~dBm. For ease of evaluation, we let each Euler angle, in \eqref{p1:Bbox} and \eqref{p1:Rbox}, of both arrays sharing the same rotation range $r$ defined by $r_{\mathrm{i},\min}^{\text{A}}=-r,r_{\mathrm{i},\max}^{\text{A}}=r,\forall \text{A}\in \left\{ \text{B},\text{R} \right\} ,\forall \mathrm{i}\in \{\mathrm{x},\mathrm{y},\mathrm{z}\}$. As the maximum allowable rotation range $r$ increases, the utility and sum rate improve (Figs.~14-15), and the NMSE decreases (Fig.~16), since a larger rotation range provides more geometric DoF to align array boresights and reshape the array response. We further observe diminishing returns: a moderate rotation capability already captures most of the gain, and further increasing $r$ yields only marginal improvements. Moreover, allowing the BS to rotate is generally more influential than rotating the RIS alone, because BS rotation directly impacts both the direct BS-user channels and the BS-RIS illumination, while RIS rotation mainly affects the reflected link. These results suggest that in practical designs, equipping the BS with moderate rotation capability and combining it with even limited RIS rotation can offer a near-optimal rate-sensing trade-off with manageable mechanical complexity.

\begin{figure}[t]
\centering
\includegraphics[width=3.2in]{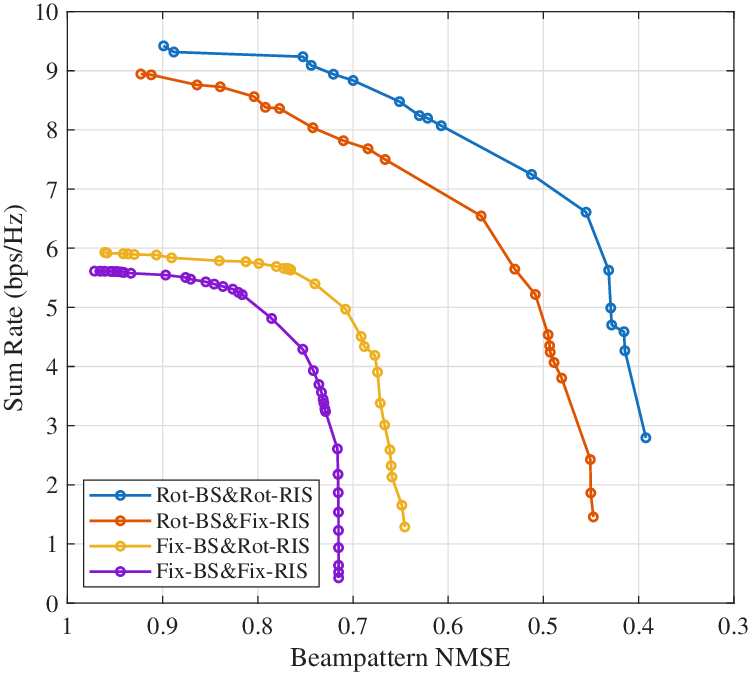}
\caption{Achievable ISAC Pareto frontier for the rate-sensing trade-off between the downlink sum rate and the transmit beampattern NMSE, obtained by sweeping the weight $\rho$ with $K$=2, $P_\text{B}$=30 dBm, $M$=2$\times$2, $N$=6$\times$6.}
\label{fig:pareto_frontier}
\vspace{-12pt}
\end{figure}

\subsection{Pareto Frontier of the Rate-Sensing trade-off}

Fig.~\ref{fig:pareto_frontier} plots the achievable rate-sensing trade-off by sweeping the weight $\rho$ in the weighted-sum objective of \eqref{p1:objective}. Each marker corresponds to the converged solution of (P1) for a given $\rho$, where we sweep $\rho$ on a logarithmic-like grid $\rho \in \{0.1, 0.2, \dots, 0.9, 1, 2,$ $ \dots,9,10,20,\dots,90,100,200,\dots, 1000 \}$.
The horizontal axis reports the beampattern NMSE in \eqref{eq:nmse_def}, where a smaller NMSE indicates a more accurate match to the desired sensing beampattern. Hence, moving rightward along each curve corresponds to increasing $\rho$ and placing a stronger emphasis on sensing fidelity, at the cost of reduced downlink sum rate.

Several key observations can be drawn. First, all schemes exhibit a clear and monotonic rate-NMSE trade-off: as $\rho$ increases, the obtained NMSE decreases, but the sum rate gradually degrades. Notably, the curves display a ``knee'' behavior, where moderate sensing improvement can be achieved with a relatively small rate loss, while pushing NMSE further down incurs a disproportionately large throughput sacrifice. This knee region offers a practical operating point for low-altitude ISAC deployment that balances sensing fidelity and multiuser throughput. Second, the fully rotatable architecture ``Rot-BS \& Rot-RIS" yields the outermost boundary. Intuitively, BS rotation provides a global geometric DoF that simultaneously impacts the direct BS-user links and the BS-side steering of the cascaded BS-RIS-user paths, thereby improving user separation and interference management.
RIS rotation further enriches the controllable spatial response of the reflected link and enables additional beampattern shaping capability, which becomes particularly valuable when stringent sensing fidelity is required.
Third, comparing the partially rotatable cases reveals the roles of active versus passive rotations. ``Rot-BS \& Fix-RIS" achieves a markedly better frontier than ``Fix-BS \& Rot-RIS", indicating that BS rotation is more influential in this setup as it directly affects all communication and sensing steering vectors. Nevertheless, enabling RIS rotation still provides a significant frontier expansion beyond ``Rot-BS \& Fix-RIS", especially in the sensing-centric regime, where the passive array can help match the desired beampattern with less distortion to the multiuser downlink beams.

\section{Conclusion}
This paper investigated a low-altitude ISAC downlink architecture that exploits mechanical rotations at both the active BS array and the passive RIS to introduce additional spatial DoF beyond conventional electronic beamforming. By incorporating rotation-dependent array geometry and element directivity into a unified directional channel model, we formulated a joint ISAC design that maximizes a weighted utility comprising the downlink sum rate and the sensing beampattern similarity, measured by NMSE. To solve the highly nonconvex problem with coupled variables, we developed an AO framework that iteratively updates the BS transmit beamforming, RIS phase shifts, and the BS and RIS array rotation angles. Specifically, the transmit beamforming subproblem is addressed by combining a QT for the sum-rate term with an MM surrogate for the sensing NMSE, yielding a structured QCQP with an efficient closed-form update under a single power constraint. The RIS phase update is performed on the complex unit-modulus manifold using RCG, while the rotation variables are updated using a projected gradient method with analytically derived Jacobians. Each block update is designed to monotonically improve the objective, and the overall procedure converges to a stationary point of the alternating iterations. Numerical results demonstrated that enabling rotations at both the BS and the RIS consistently enlarges the achievable ISAC performance region compared with fixed-array or partially rotatable baselines, with the gains becoming more pronounced under practical directional element patterns. Moreover, by sweeping the utility weight, we obtained achievable rate-sensing Pareto frontiers that explicitly quantify how mechanical rotations expand the trade-off boundary. 

Several important extensions remain for future work. First, incorporating imperfect CSI and sensing-target uncertainty would enable robust designs aligned with practical estimation errors. Second, the time-scale mismatch and actuation overhead of mechanical rotations (e.g., latency, energy consumption, and motion constraints) should be modeled to capture dynamic operation. Finally, extending the framework to wideband channels, multi-target sensing, and discrete or quantized RIS hardware constraints would further strengthen the applicability of rotatable active-passive arrays in realistic low-altitude ISAC systems.

\ifCLASSOPTIONcaptionsoff
  \newpage
\fi

\end{document}